\documentclass[twocolumn]{aastex62}

\usepackage{graphicx}
\usepackage{amsmath}
\usepackage{hyperref}

\usepackage{color}

\graphicspath{{./}{figures/}}

\received{January 1, 2018}
\revised{January 7, 2018}
\accepted{\today}
\submitjournal{ApJ}

\DeclareUnicodeCharacter{2212}{-}

\shorttitle{Kinematics in 3C 84}
\shortauthors{Hodgson et al.}

\begin{document}

\title{A detailed kinematic study of 3C\,84 and its connection to $\gamma$-rays}

\correspondingauthor{Jeffrey A. Hodgson}
\email{jhodgson@sejong.ac.kr}

\author[0000-0001-6094-9291]{Jeffrey A. Hodgson}
\affil{Department of Physics and Astronomy, Sejong University, 209 Neungdong-ro, Gwangjin-gu, Seoul, South Korea}
\affil{Korea Astronomy and Space Science Institute
Daedeokdae-ro 776
Daejeon, South Korea}

\author{Bindu Rani}
\affil{Korea Astronomy and Space Science Institute
Daedeokdae-ro 776
Daejeon, South Korea}
\affil{NASA Goddard Space Flight Center, Greenbelt, MD, 20771, USA }
\affil{Department of Physics, American University, Washington, DC 20016, USA}

\author{Junghwan Oh}
\affil{Korea Astronomy and Space Science Institute
Daedeokdae-ro 776
Daejeon, South Korea}
\affil{Department of Physics and Astronomy, Seoul National University, 1 Gwanak-ro, Gwanak-gu, Seoul 08826, Korea }

\author{Alan Marscher}
\affil{Institute for Astrophysical Research, Boston University, 725 Commonwealth Avenue, Boston, MA 02215}

\author{Svetlana Jorstad}
\affil{Institute for Astrophysical Research, Boston University, 725 Commonwealth Avenue, Boston, MA 02215}
\affil{Astronomical Institute, St. Petersburg State University, Universitetskij Pr. 28, Petrodvorets,
198504 St. Petersburg, Russia}

\author{Yosuke Mizuno}
\affil{Tsung-Dao Lee Institute and School of Physics and Astronomy, Shanghai Jiao Tong University, 800 Dongchuan Road, Shanghai, 200240, People’s Republic of China}
\affil{Institut f\"{u}r Theoretische Physik, Goethe Universit\"{a}t, Max-von-Laue-Str. 1, 60438, Frankfurt am Main, Germany }

\author{Jongho Park}
\affil{Institute of Astronomy and Astrophysics, Academia Sinica, P.O. Box 23-141, Taipei 10617, Taiwan}

\author{S.S. Lee}
\affil{Korea Astronomy and Space Science Institute
Daedeokdae-ro 776
Daejeon, South Korea}
\affil{University of Science and Technology, Gajeong-ro, Yuseong-gu, Daejeon, South Korea}

\author{Sascha Trippe}
\affil{Department of Physics and Astronomy, Seoul National University, 1 Gwanak-ro, Gwanak-gu, Seoul 08826, Korea}

\author{Florent Mertens}
\affil{Kapteyn Astronomical Institute, University of Groningen, P. O. Box 800, 9700 AV Groningen, The Netherlands}
\affil{Max-Planck-Institut f\"{u}r Radioastronomie, Auf dem H\"{u}gel 69, 53121, Bonn, Germany }





\begin{abstract}
3C\,84 (NGC\,1275) is the bright radio core of the Perseus Cluster. Even in the absence of strong relativistic effects, the source has been detected at $\gamma$-rays up to TeV energies. Despite its intensive study, the physical processes responsible for the high-energy emission in the source remain unanswered. We present a detailed kinematics study of the source and its connection to $\gamma$-ray emission.
The sub-parsec scale radio structure is dominated by slow moving  features in both eastern and western lanes of the jet. The jet appears to have accelerated to its maximum speed within less than 125 000 gravitational radii. The fastest reliably detected speed in the jet was $\sim$0.9\,c. This leads to a minimum Lorentz factor of $\sim$1.35. Our analysis suggests the presence of multiple high-energy sites in the source.  If $\gamma$-rays are associated with kinematic changes in the jet, they are being produced in both eastern and western lanes in the jet. Three $\gamma$-ray flares are contemporaneous with epochs where the slowly moving emission region splits into two sub-regions. We estimate the significance of these events being associated as $\sim 2-3\sigma$. We tested our results against theoretical predictions for magnetic reconnection induced mini-jets and turbulence and find them compatible.
%

\end{abstract}



\section{Introduction}


Gamma-ray bright misaligned  radio galaxies
provide a unique opportunity to probe the high-energy emitting sites and particle acceleration processes \citep{rani2019}. Being  seen off-axis, one could transversely resolve the fine scale structure of the jet and study its connection with $\gamma$-ray emission. We present here the jet kinematics study of a nearby misaligned radio galaxy 3C~84  \citep[z=0.017559, ][]{3c84_z} and its connection with the $\gamma$-ray flaring activity. The source is the radio counterpart of the Seyfert Type 1.5 galaxy NGC\,1275 (also known as Perseus A) \citep{ho97}. The {\it Fermi}-Large Area Telescope (LAT) has been detecting  GeV emission from the source since the beginning of its operation in 2008 \citep{fermi2009}. \citet{TeV_detection} confirmed  the detection of TeV emission ($>$100~GeV)  from the source, followed by several extremely bright and exceptionally rapid TeV flares detected in 2016-2017 \citep{3c84_atel16_1, 3c84_atel16_2, 3c84_atel16_3, 3c84_atel16_4, 3c84_atel16_5}.

Several attempts have been made to understand the $\gamma$-ray flux and spectral variability and its connection to broadband emission in 3C~84 \citep[][and references therein]{TeV_detection, magic2018, fermi2009, hodgson18}.
A problem in high energy astrophysics is the ``Doppler crisis'', where there are theoretical difficulties in reproducing the observed GeV and TeV photons with very fast variability time-scales in the misaligned sources without high Doppler factors to boost the emission \citep[e.g.][]{albert07}. This has been reported to be an issue in 3C\,84. It has been proposed that the high energy emission could be explained by emission from the magnetospheric gap around the central SMBH \citep{magic_3c84_18}. Another potential explanation could be having the emission generated near the leading edges of ejected jet components \citep{lyutikov10} and finally, high Doppler factors could be generated by via magnetic reconnection powered ``mini-jets'' \citep{giannios09,giannios13}.

\begin{figure}\label{stacked}
\center
\includegraphics[scale=0.6, trim=410 100 410 10, clip=true]{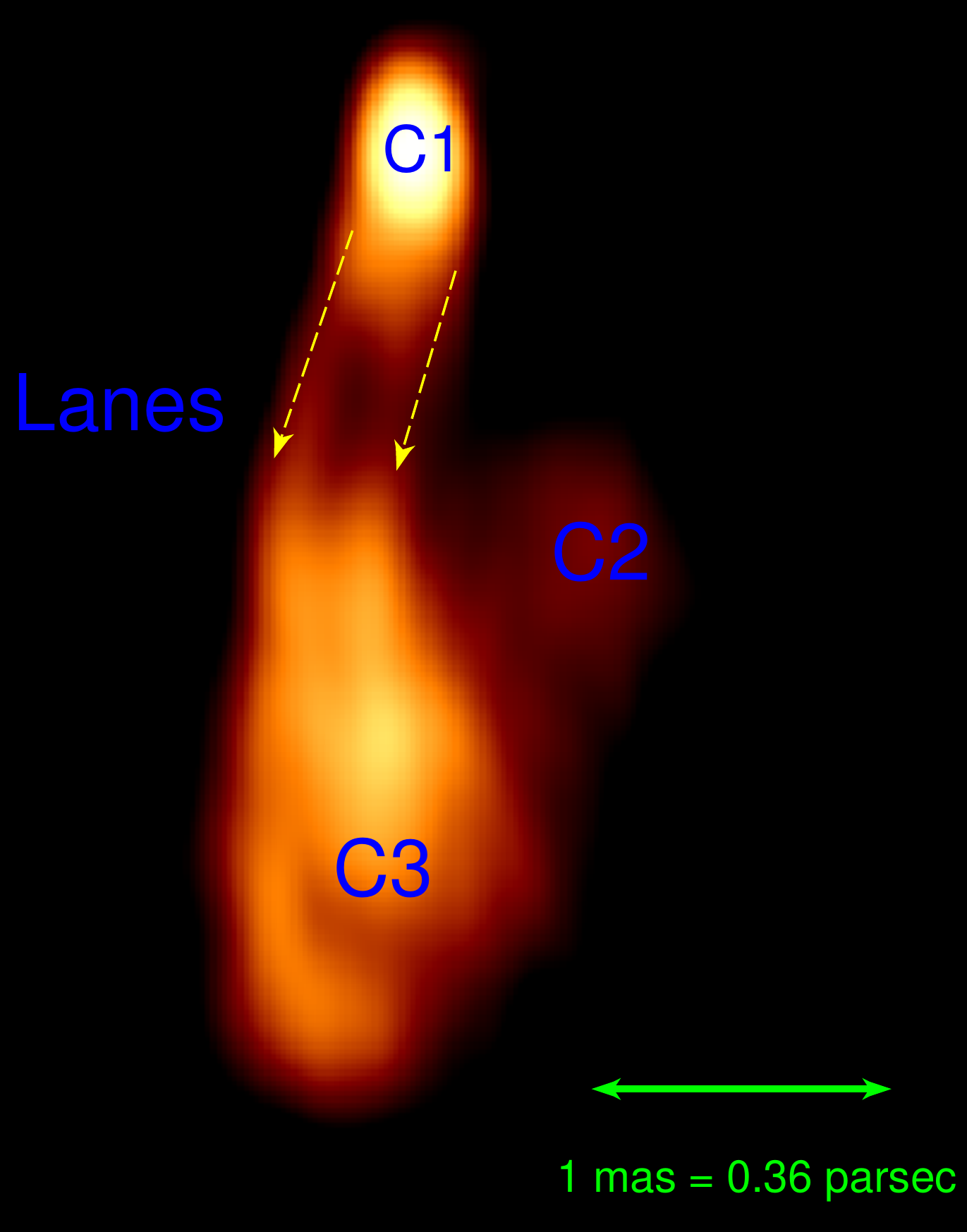}
\caption{Stacked 43 GHz image of 3C\,84, with the major emission regions labeled. Motion detected in the two lanes is found to be potentially associated with $\gamma$-ray flares (see Section 3 for details). }
\end{figure}

Being one of the brightest radio galaxies, the jet morphology of the source has been extensively studied.
Early studies speculated that the radio emission was due to colliding galaxies \citep[e.g.][]{baade_win_54}. However, further observations suggested that the radio emission was in fact due to highly energetic and fast moving emission from the nuclear region of the galaxy \citep{BBS63,BB65}. The relationship between the colliding galaxies and the AGN activity is still unclear  \citep{rubin77, haschick82, hu83}. The radio emission of 3C\,84 had been declining since the 1950's.
The most recent jet activity appears to have started in $\sim$2005, coinciding with a general increase in flux density in the source and creating the region known as ``C3", which was ejected from the presumed jet launching region ``C1'' (see Fig.~1). Additionally, there is a large, faint area of quasi-stationary emission known as ``C2'' that is $\sim$40$^{\circ}$ offset from the current jet emission (see: Fig. \ref{stacked}), and is presumed to be related to previous jet activity \citep{nagai10}.

Recent   space-VLBI observations with RadioAstron revealed a very broad jet base and that the mas-scale cylindrical jet extends to within a few hundred $R_{\mathrm{g}}$ of the central SMBH \citep{radio_astron_3c84}. At a similar angular resolution, \citet{kim19} using global 3\,mm VLBI, reported on the polarization properties of the core region, finding inhomogeneous and mildly relativistic plasma. On more extended scales, the source exhibits non linear motions within the C3 component \citet{hiura18}. 3C\,84 also exhibits a counter-jet \citep{vermeulen94}, which has recently been used to determine an inclination angle of $\sim65^{\circ}$ to the line-of-sight by \citet{fujita17}, which is in tension with the $\sim 25^{\circ}$ inclination angle determined from $\gamma$-ray analysis \citep{abdo09}. Most recently, \citet{kino18} has analysed VLBI observations (including some that is also used in this study) of the C3 region, finding a ``jet-flip'', where the direction of the jet appeared to drastically change. They interpreted this as being due to the interaction of C3 with the external medium, consistent with the results of \citet{nagai17}, where further evidence was found that the 3C\,84 jet is likely in a clumpy external medium. These interactions have been proposed to be where at least some of the $\gamma$-ray emission could be occurring \citep{nagai14,nagai17}. The parsec scale environment of 3C 84 was analyzed by \citet{wajima2020}, who suggested that 3C\,84 could be surrounded by either an assembly of clumpy clouds or a dense ionized gas in these regions.

The mm-wave radio and $\gamma$-ray emission was recently analyzed by \citet{hodgson18} (hereafter referred to as Paper I, with the $\gamma$-ray light-curve reproduced here in Fig. \ref{G_lc} for the benefit of the reader, with the most significant $\gamma$-ray and mm-wave radio flares reproduced in Table \ref{gamma_flares} and Table \ref{sma_flares} respectively.), using multi-wavelength Korean VLBI Network (KVN) data combined with data from the Fermi Large area telescope (\emph{Fermi}-LAT). Our study suggests that the long-duration $\gamma$-ray flares are most likely produced in the C3 region, while the rapid variations seems to be occurring in the region closer to the central black hole (C1).

In this paper, we present the results of the wavelet-based image segmentation and evaluation (WISE) analysis method for the jet kinematic analysis of 3C\,84 using 7\,mm VLBA data from 2010 until 2017, and have compared it against CLEAN maps.
The paper is structured as follows. Section 2 presents observations and data analysis. Results are given in Section 3, and discussed in Section 4. Section 5 summarizes our main findings.
The distance to the source was recently estimated by \citet{hodgson20}, using VLBI and variability and the speed of light to calibrate the source as a standard ruler, however the systematic uncertainties related to the method are still not well known. We therefore use a flat $\Lambda$CDM cosmology with $H_{0}$=69.6\,$\mathrm{km/s/Mpc}$ and $\Omega_{\rm m}$=0.286 \citep{bennett14}, corresponding to a linear scale of 1\,mas = 0.359\,pc at a luminosity distance of $D_{\rm L}$=74\,Mpc. There is quite a large range in published values for the BH mass, from 3$\times 10^{7}$ to 8$\times 10^{8}$ M$_{\odot}$ and as high as 1$\times 10^{9}$ M$_{\odot}$ \citep{nagai19}. We will discuss the issues regarding the black hole mass in a future paper, but here we assume a mass of $M_{\rm BH} \sim 3 \times 10^{7} M_{\odot}$ \citep{onori17}, which corresponds to projected scales of $\sim$350,000\,$R_{\rm S}$/pc or $\sim 125 000~R_{\rm S}$/mas. If the mass is larger than this, the projected distances will be relatively closer to the SMBH in units of $R_{\rm G}$.

\begin{figure*}
\includegraphics[width=\linewidth]{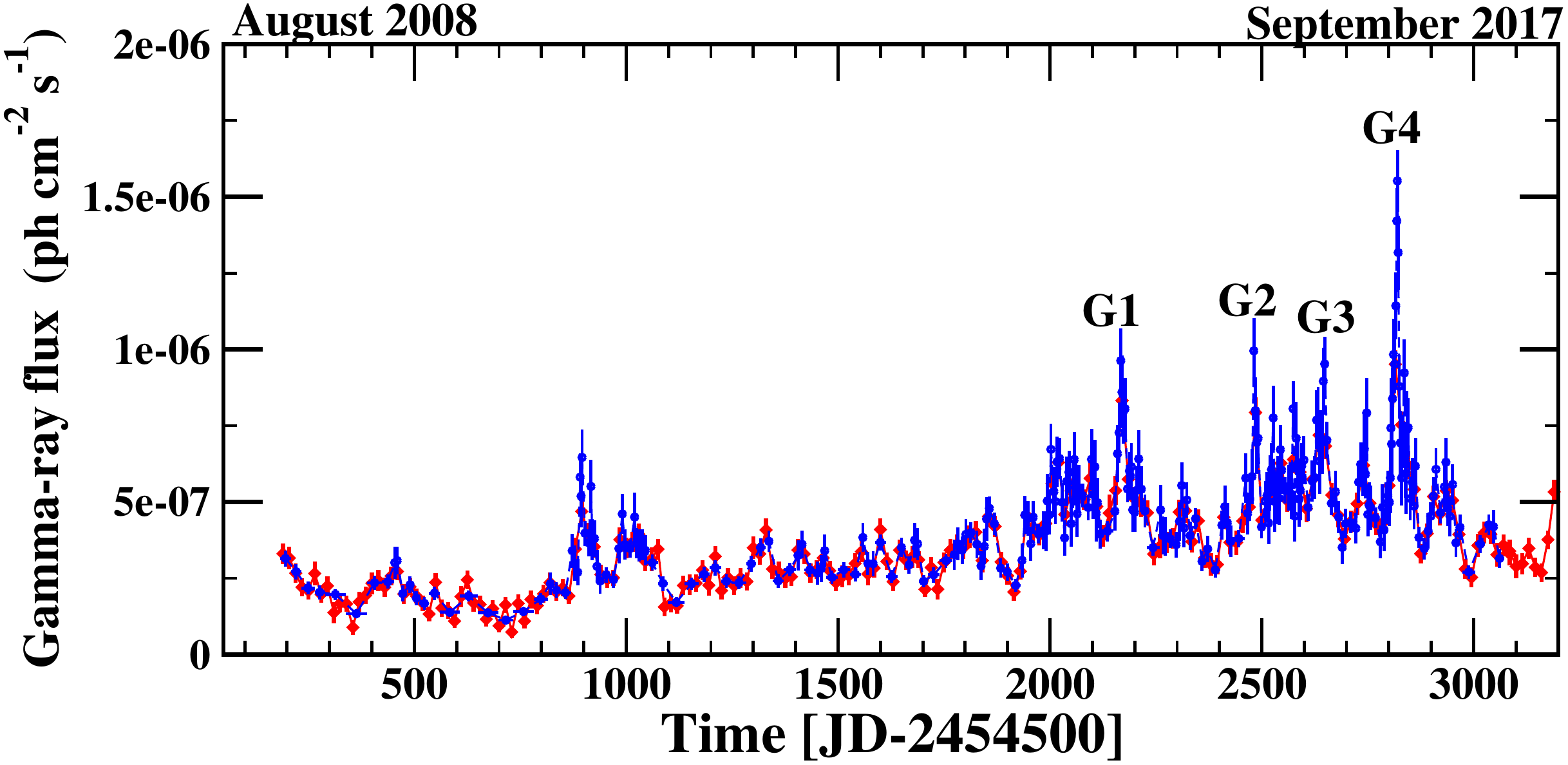}
\caption{Adaptive binned (15$\%$ uncertainty in blue) and 15-day binned (in red)  $\gamma$-ray photon flux light curves of 3C~84, reproduced from Paper I. The most significant $\gamma$-ray flares are labelled as G1 to G4. Note that many low-amplitude flares have also been observed in between these events. }
\end{figure*}\label{G_lc}



\begin{figure*}
\includegraphics[width=\textwidth]{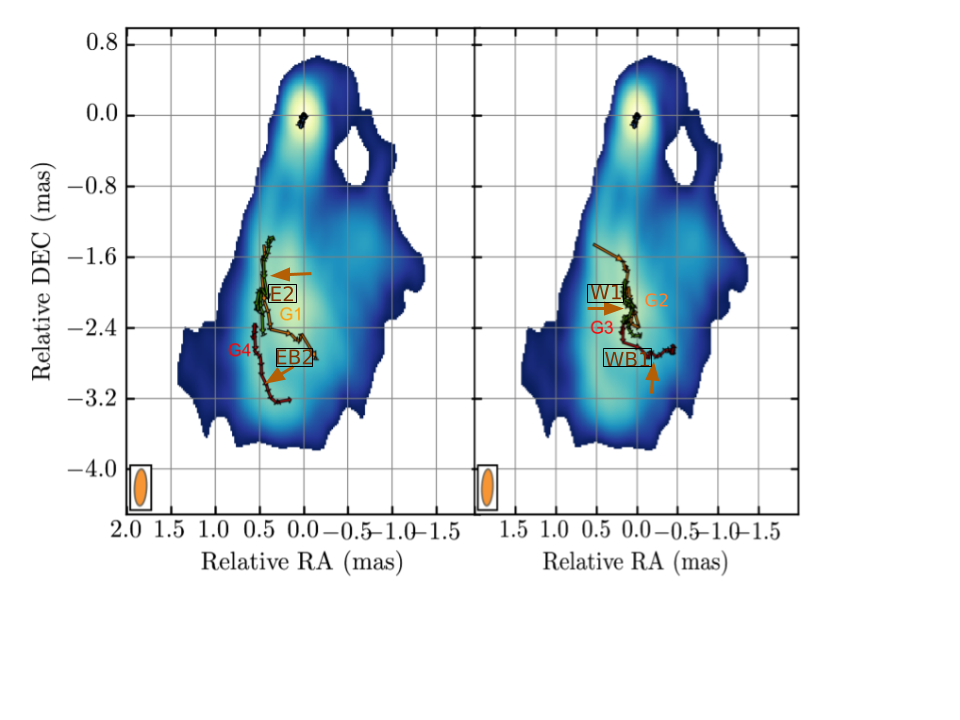}
\caption{Consolidated component trajectories detected in 3C 84 (from Table \ref{kinematics}). Eastern lane is in the left panel, while the western lane is shown in right panel. Component names are bordered, while flare labels are not. The flare names are the approximate location of where potential $\gamma$-ray flare associations have been made. } \label{xy}
\end{figure*}



\section{Observations and data reduction}

A total of 52 VLBI maps were obtained from the VLBA-BU-BLAZAR monitoring program from 2010.87 until 2017.03. The program monitors $\gamma$-ray bright blazars and radio galaxies at 43\,GHz (7\,mm) with a roughly monthly cadence using the Very Long Baseline Array (VLBA). See \citet{jor05} and \citet{jor17} for details of data reduction and results from the program. {\textit {u,v}} data and CLEAN model files were downloaded from their website and analysed using DIFMAP \citep{difmap}. The maps were then re-gridded to a common map and pixel-size and then aligned on the brightest, northernmost component (C1). The data were calibrated in dual polarization (LCP and RCP), thus allowing for polarization maps to be produced, although in this paper we investigate only the total intensity emission. \\

Because of the extended jet morphology and limited VLBI resolution, we miss the extended jet emission for 3C 84, which affects its amplitude calibration \citep[for more details, see][]{hodgson20}. In this paper, we concentrate on any apparent correlations with the kinematic behavior and $\gamma$-ray emission, which is not sensitive to the amplitude calibration.

\section{Results}



\subsection{Time variable morphology}\label{time_var_morph}

Viewing the maps in a sequence (with a link to a movie in the footnote\footnote{Images can be viewed at \url{https://www.bu.edu/blazars/VLBA_GLAST/0316.html} or an animated gif can be found at \url{https://www.dropbox.com/s/ep6bc9b7d0a4mb9/3c84_movie.gif?dl=0}}, or can be viewed on the BU website), beginning in late 2010, a bright hotspot (which is sometimes also referred to as a knot in the literature) is seen south of C1, separated by about 1.5\,mas, but appears to be behind the shock front. In 2013, a hotspot appears in western lane at the the leading edge of the emission region, becoming much brighter and moves further to the west. During the period of becoming brighter, the component sometimes appears to split into two components before dissipating. As the western hotspot becomes fainter, beginning in late 2016, in the eastern lane another hotspot also at the leading edge of the high intensity ridge appears and follows a similar western trajectory. There then appears to be some trailing emission in the direction of C2. Viewed as a whole, it takes the appearance of some material moving away from C1 and hitting the C2 region.

\subsection{WISE Analysis}\label{sec:wise_analysis}

We apply the WISE analysis package to the VLBI data to derive kinematics at various scales.
The wavelet-based image segmentation and evaluation (WISE) method combined with a stacked cross-correlation algorithm \citep{mertlov16}, was developed by \citet{mertlov15}. Traditionally, the kinematics of a source were determined by first characterizing emission regions within a VLBI map by fitting Gaussian models to the visibilities over multiple epochs. Emission regions identified over multiple epochs would then be cross-identified by eye and properties such as proper motions derived. In comparison, WISE allows for the \emph{objective} determination of these cross-identifications. It should, however, be emphasized that the results of the algorithm may not necessarily represent the true two-dimensional velocity field. The technique was applied to the nearby (and often compared with 3C\,84) AGN M\,87 \citep{mertlovm87}.
A more detailed description of the kinematic analysis can be found in \citet{mertlov15}.
The algorithm works by decomposing and segmenting each map and then each consecutive pair of maps is cross-correlated in order to determine the significance of the correlation and thus likelihood that a segment is related from map to map. We set a significance threshold of 3$\sigma$ for segments in a map and a correlation threshold of 0.6. The errors are determined using Eq. 1 in \citet{mertlovm87}. Errors are sometimes less than 0.01 mas/year, particularly on components such as the core. Sometimes, negative motions are detected, but these are likely anomalies caused by inaccuracies in the WISE analysis. We can vary the resolution of the decomposition and segmentation from approximately the size of the minor axis of the beam to slightly larger than the major axis of the beam, which we call the ``scale''. These scales are analogous to convolving the maps with smaller or larger beams respectively. As we go to smaller scales, we are more likely to have spurious apparent systematic motions.

When a component is cross-identified in consecutive epochs, it is called a chain. In general, the longer the chain, the more confident we are in the detection. For this reason, we set a minimum chain length of cross-identified components from the WISE analysis.
In effect, this means that we are disregarding fits that are not found in more than some threshold number of consecutive epochs. Therefore, a ``4 component fit'' would mean fits that are found in at least four consecutive epochs. An ``8 component fit'' means a fit found in at least eight consecutive epochs, with the 8-link chains becoming more important as we analyze at smaller scales. This is because it will increase the significance of the detected motions. We began with the larger scale analysis and worked towards higher resolutions. In our analysis, we only analyzed the region south of the C1 region. We will analyze the counter-jet in a future paper. We then composed Table \ref{kinematics}, which consists of consolidated motions by combining the motions of components that were similar for different resolutions. The full results of the analysis are shown at the end of the paper in Tables \ref{wise_048_8} to \ref{wise_016_4}. Components are named first by the scale and then a number automatically assigned by the algorithm. For example a component labeled 5 by the algorithm and detected at the 0.48\,mas scale would be named \emph{24:5}. In general, the effect of reducing the WISE analysis scale is to increase the kinematic resolution, but at the expense of more scatter in the component motions detected.
%

\subsubsection{0.48 mas scale}

This scale (Fig. \ref{048mas8} - Fig. \ref{048mas4} and Table \ref{wise_048_8} - Table \ref{wise_048_4}) gives the best fits with least amount of scatter. There are no significant differences between the 4 and 8 component fits. Component 24:9, south of the brighter parts of the C3 region emerges in late 2015 at 0.65c. At a similar time, component 24:1 moves rapidly to the west. No motion or very slow inward motion is found in the C1 region. C2 is quasi-stationary (e.g. component 24:5 at 0.14c).

\subsubsection{0.32 mas scale}

We now reduce the WISE analysis scale to be the average major axis of the beam (Fig. \ref{032mas8} - Fig. \ref{032mas4} and Table \ref{wise_032_8} - Table \ref{wise_032_4}). We see broadly consistent results with the 0.48mas scale, component 16:17 is the same as component 24:9, with the same speed of 0.47c. Component 16:1 is the same as 24:1, with a similar speed. This component is still detected until the most recent data, continuing on a westerly trajectory. With 4 component fits, some motion is detected in the ``lane'' area, with for example component 16:18 having a speed of 0.44c.

\subsubsection{0.24 mas scale}

This length scale is between the major and minor beam axes, with the results of the analysis in Fig. \ref{024mas8} - Fig. \ref{024mas4} and Table \ref{wise_024_8} - Table \ref{wise_024_4}. The lanes are clearly detected separately at this scale. Component 12:11 (0.66c) follows an apparently curved path, starting on the western lane, crossing to the eastern lane in mid-to-late 2012. It then crosses back from the eastern lane to the western lane in early 2014. This coincides with the previous components 16:1 and 24:1, which were one slow moving component, splitting in two. Now these components are identified as 12:1 until early 2014. Two components were then identified, with component 12:47 being likely the same as 16:17 and 24:9. Another component is detected in the western lane (12:46) with a relatively fast motion of 0.66c although having started in the eastern lane, it then moved across the jet in early-to-mid 2015. In early 2016, an apparently quasi-stationary feature has formed (component 12:61) at a separation of $\sim$2\,mas from C1. Additionally in the 4 component detection scheme, as with the 0.32\,mas scale analysis, some motions are detected in the ``laned'' region with similar speeds.


\subsubsection{0.16\,mas scale}

The analysis here is performed on the scale of the minor axis of the beam and is likely close to the limit of what could be assumed to be a reliable analysis (Fig. \ref{016mas12} - Fig. \ref{016mas4} and Table \ref{wise_016_12} - Table \ref{wise_016_12}). In addition to the 4 and 8 component fits, we have included a 12 component fit in order to increase our confidence in the results at this scale. In general, it is consistent with the results seen in the 0.24\,mas scale analysis. The most striking difference is the detection of an additional component in the eastern lane (component 8:8, 0.47c), which although travelling almost the same trajectory as component 8:31 (cross-identified with component 12:11), is considerably slower than component 8:31's 0.89c. It additionally does not appear to cross from the eastern to western lane. In the 12 component fit, the westerly motion of components 12:36 and 16:1 are not detected although in the 8 component fit it is seen as component 8:78. In the 4 component fit plots we note that almost all components at a separation of $\sim$0.4-1.2\,mas appear to follow a curved trajectory moving from the west to the south-east, although interpreting these should be done with caution, as 4-component fits at the smallest length scale tend to be the most unreliable.

\begin{figure}
\includegraphics[width=\linewidth]{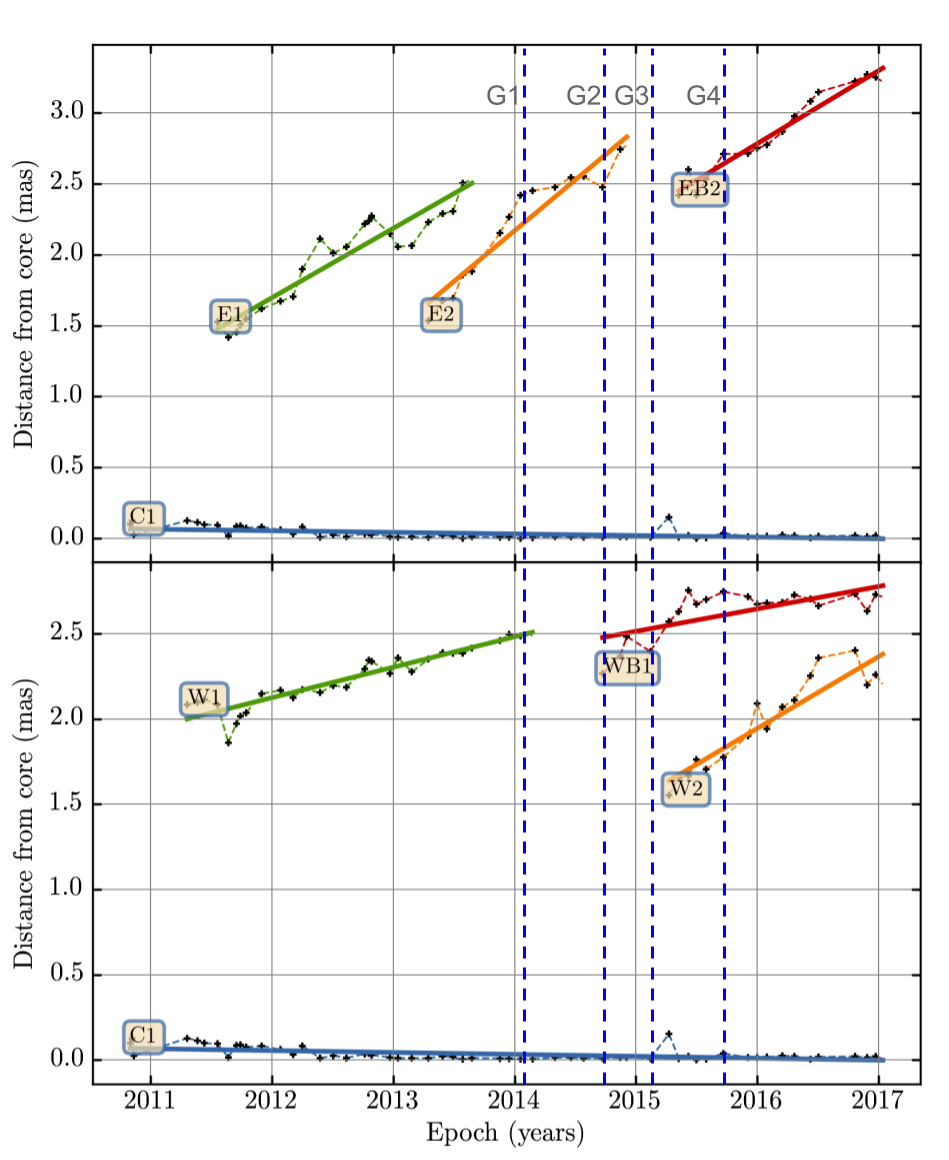}
\caption{Consolidated component motions, as determined from the multi-scale WISE analysis. Eastern lane (top) and western lane (bottom). }\label{core_sep}
\end{figure}

\begin{figure*}\centering
\includegraphics[width=0.75\textwidth]{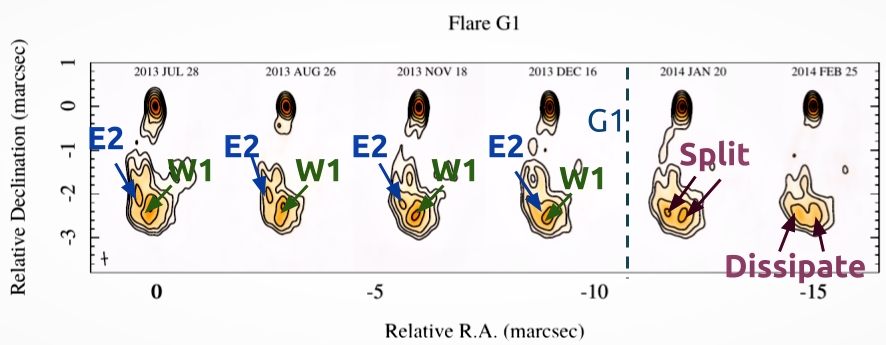}
\caption{CLEAN maps over the period of the flare G1. A $\gamma$-ray flare occurs between the 2013.96 (2013 Dec 16) and 2014.05 (2014 Jan 20) epochs. } \label{G1_maps}
\end{figure*}

\begin{figure*}\centering
\includegraphics[width=0.75\textwidth]{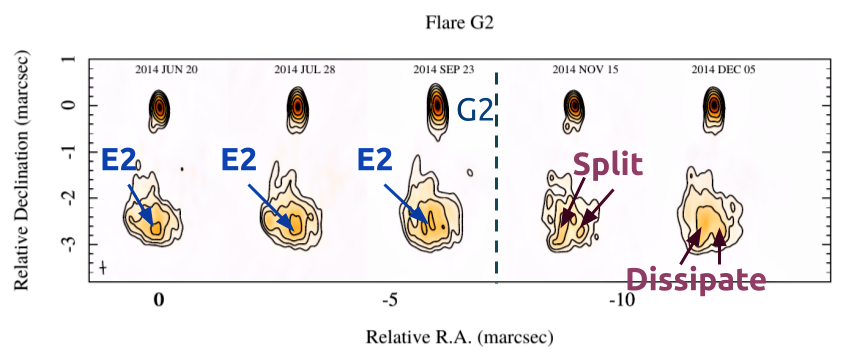}
\caption{CLEAN maps over the period of the flare G2. A $\gamma$-ray flare occurs between the 2014.73 (2014 Sep 23) and 2014.87 (2014 Nov 15) epochs. } \label{G2_maps}
\end{figure*}

\begin{figure*}\centering
\includegraphics[width=0.75\textwidth]{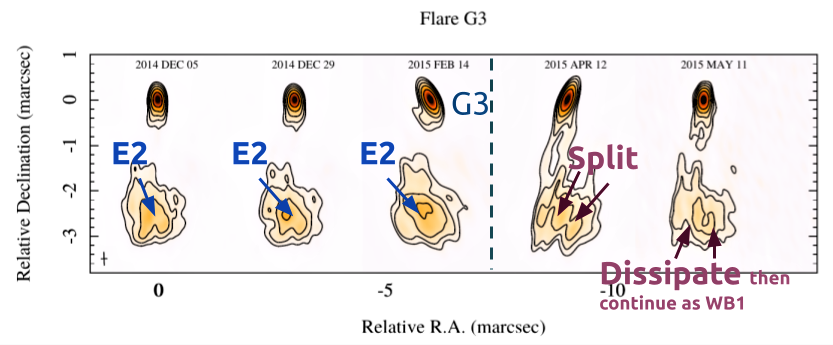}
\caption{CLEAN maps over the period of the flare G3. A $\gamma$-ray flare occurs between the 2015.12 (2015 Feb 14) and 2015.28 (2015 Apr 12) epochs. } \label{G3_maps}
\end{figure*}

\begin{figure*}\centering
\includegraphics[width=0.75\textwidth]{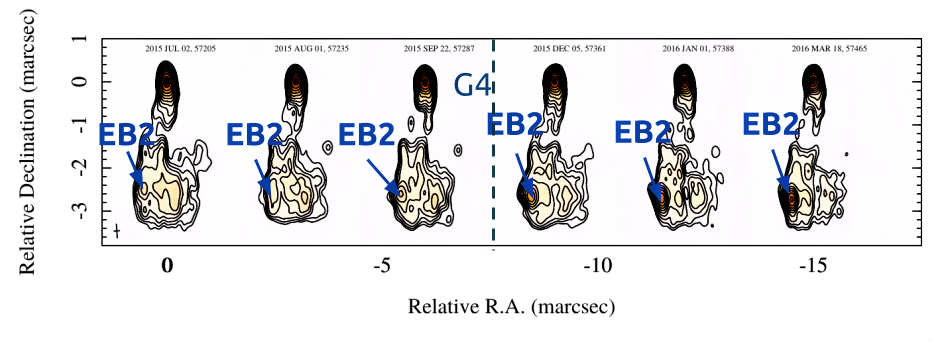}
\caption{CLEAN maps over the period of the flare G4. A $\gamma$-ray flare occurs between the 2015.72 (2015 Sep 22) and 2015.93 (2015 Dec 05) epochs. } \label{G4_maps}
\end{figure*}

\begin{table}
	\centering
	\caption{Significant $\gamma$-ray flares since 2013 as determined from Paper I.}
	\label{gamma_flares}
	\begin{tabular}{lccc} 
		\hline
		ID & Date & Flux density & Spectral Index\\
		   &      & [x10$^{-7}$] ph cm$^{-2}$ s${-1}$ & \\
		\hline
		G1 & 2014.04 & 10.61 $\pm$ 1.28 & 2.16 $\pm$ 0.10 \\
		G2 & 2014.88 &  9.67 $\pm$ 1.26 & 2.15 $\pm$ 0.10 \\
		G3 & 2015.14 &  9.47 $\pm$ 1.17 & 2.27 $\pm$ 0.10 \\
		G4 & 2015.81 & 13.45 $\pm$ 1.27 & 1.97 $\pm$ 0.06 \\
		\hline
	\end{tabular}
\end{table}

\begin{table}
	\centering
	\caption{Significant 1\,mm radio flares since 2013 from Paper I.}
	\label{sma_flares}
	\begin{tabular}{lccc} 
		\hline
		ID & Peak of the flare & Duration     & Flux density \\
		   &      &  [months] & [Jy]  \\
		\hline
		R1 & 2013.14  &  $\sim$5    & 11.19 $\pm$ 0.56 \\
        R2 & 2014.61 & $\sim$3      & 12.67 $\pm$ 0.65 \\
        R3 & 2016.65 & $>$14  & 17.11 $\pm$ 0.85 \\
		\hline
	\end{tabular}
\end{table}

\begin{table*}
	\centering
	\caption{Consolidated component motions. Components from the multi-scale WISE analysis with similar trajectories were cross identified and consolidated. The values ``a:b'' are the component identifications automatically assigned by the WISE algorithm.  }
	\label{kinematics}
	\begin{tabular}{lcccccc} 
		\hline
		ID & 0.16 & 0.24  & 0.32 & 0.48 & $\beta_{\rm app}$ & First identified \\
		   &      &       &      &      &   [c]         &    \\
		\hline
		S &      &       & 16:1 & 24:1 &  0.24 (0.01)  & - \\
		Sa &      & 12:1 &  -   &  -   & 0.34 (0.01)   & 2011.3 \\
		W1 & 8:9  & -     &  -   &  -   & 0.23 (0.01)   & 2011.3 \\
		E1 & 8:8  & -     &  -   &  -   & 0.47 (0.01)   & 2011.3 \\
		E2 & 8:31 & 12:11 &  -   &  -   & 0.89 (0.05)   & 2013.2 \\
		W2 & 8:66 & 12:46 &  -   &  -   & 0.66 (0.05)    & 2015.3 \\
		WB1 & 8:63,8:78 & 12:36 & -    & -    &  0.21 (0.02)   & 2014.9 \\
		EB2 & 8:68 & 12:47 & 16:17 & 24:9 & 0.65 (0.03)  & 2015.5 \\

		\hline
	\end{tabular}
\end{table*}




\subsection{Flaring activity association with jet kinematics}\label{flare_associations}
In this section, we investigated the association between the observed $\gamma$-ray flaring activity and jet morphology.
In Table \ref{gamma_flares} and Table \ref{sma_flares}, we have reproduced the results concerning the radio and $\gamma$-ray flare peaks determined in Paper I. In Table \ref{kinematics}, we have combined component motions that could be cross-identified across different analysis resolutions and therefore considered to be the most relevant. We have also included the average apparent component speed and the approximate epoch at which the component appeared. For the aid of the reader, we have plotted the the consolidated trajectories in Fig. \ref{xy} and the consolidated core-separation plot in Fig. \ref{core_sep}, which we have labeled with the components found to be potentially associated with $\gamma$-ray activity. If a structural change was observed in the VLBI maps, we considered $\gamma$-ray activity to be associated if it occurred within the time between the previous and the following epochs.


The  component E2 showed non-linear motion averaged over by the WISE algorithm. The component at $\sim$2014.1 suddenly changes trajectory, moving westwards, having previously been moving in a southerly direction. We also see a change in the jet speed, from 0.9 to 1.1~$c$, at this time as shown in Fig. \ref{core_sep}.
When component E2 changes to its westerly trajectory and appears to interact with W1, it almost exactly coincides with the $\gamma$-ray flare G1. Flares G3 and G4 appear to coincide with the appearance of the westerly trajectories of components WB1 and EB2 respectively.

\subsubsection{In-depth analysis of the potential associations}

From the WISE analysis described in Section \ref{sec:wise_analysis}, we have found that the complex kinematic behavior could potentially be associated with $\gamma$-ray activity. In order to further explore the association, we have presented CLEAN maps, concentrating on the periods around each of the $\gamma$-ray flares. In Fig. \ref{G1_maps}, we found a fast moving component E2, that was moving southerly along the eastern lane. It then began to interact with the component W1 that had previously been traveling along the western lane. Beginning late 2013, the region containing both components becomes brighter with both components having comparable brightness, appearing almost as a single large emission region by December 2013. Flare G1 is observed  before the component split into two separate regions and then dissipated quickly. A single region then brightens again, with this component presumed to be component E2 from the WISE analysis. The component E2 is either detected as a single continuous component over the period from $\sim$G1 until G3, or detected as several components, depending on the length scale of the WISE analysis. We have kept the E2 designation from the larger length scale for the benefit of the reader.  \\

In Fig. \ref{G2_maps}, component E2 continues to move on a westerly trajectory, consistent with the WISE analysis, becoming brighter and then in the VLBI images immediately after the $\gamma$-ray flare G2, (between the 2014 Sep 23 and 2014 Nov 15 maps), the region again is resolved into two separate emission regions, which then  dissipate. A similar behavior can be seen in Fig. \ref{G3_maps}, where component E2 continues moving easterly, expanding and becoming brighter before splitting into two components. The ``flare-split-dissipate'' behavior appears to be common across flares G1, G2 and G3. In the later two cases however, the components continue as EB2 and WB1, respectively.  

We noticed a new emission region/component, EB2, in the western lane of the VLBI images at the time of flare G4 (see Fig. \ref{G4_maps}). Then EB2 continues to brighten, producing the radio flare R3, which correlates with the $\gamma$-ray flare G4 (see Paper I for the detailed light curve analysis). This behavior is different from the previous three flares and appears to be coincident with the onset of radio flare R3. Also note that this flaring activity appears to be happening in the western lane, while flares G1 to G3 seem to be associated with the jet activity in the eastern lane.

\subsubsection{Significance of kinematic associations}

\begin{figure}
\includegraphics[width=0.99\linewidth]{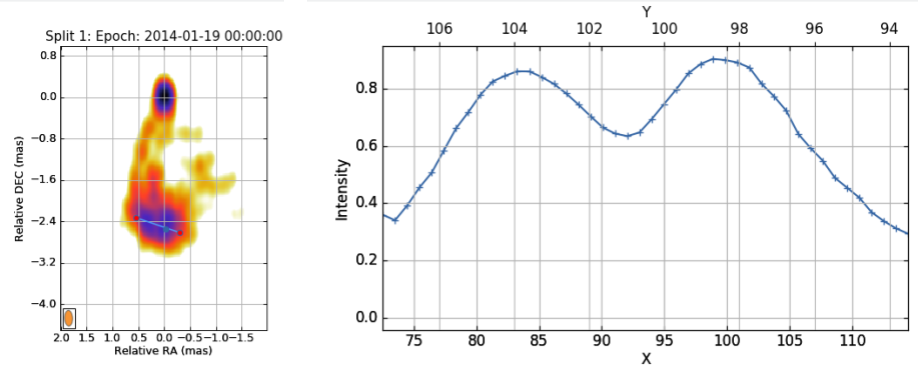}
\includegraphics[width=0.99\linewidth]{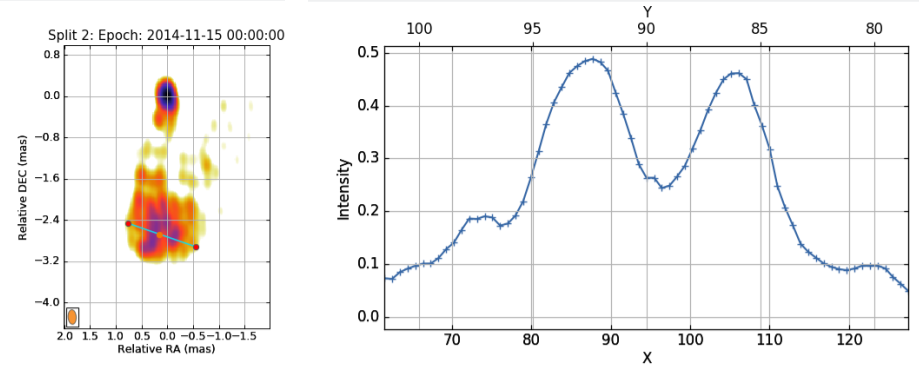}
\includegraphics[width=0.99\linewidth]{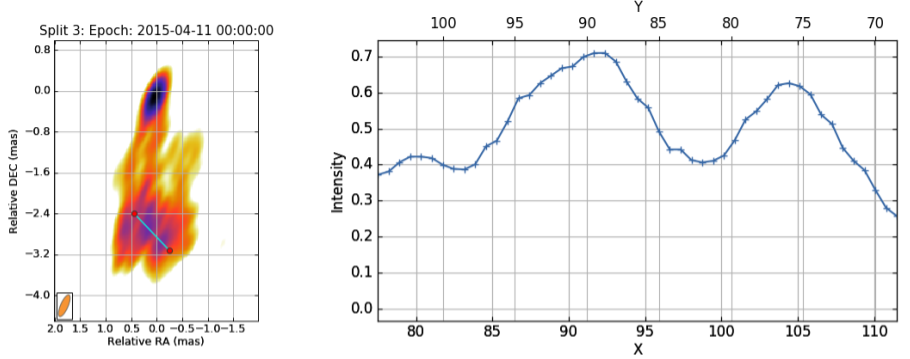}
\caption{Ridge-line analysis of the three ``split'' epochs. Intensity if normalized to the maximum flux density.}  \label{ridge_lines}
\end{figure}

In this paper, we have presented evidence for a potential connection between the $\gamma$-ray flaring and the kinematic behavior in 3C\,84. In order to estimate the significance of these associations, we first determine the significance of individual ``split-flare-dissipate'' events and then determine the likelihood that these events are randomly associated with $\gamma$-ray flaring.

To determine the significance of individual ``split-flare-dissipate'' events, we first investigated the ridge-line  profiles of the region (see  Fig.\ \ref{ridge_lines}). We find that in all three cases, the ``split'' appears significant. To further investigate, we then compared single and double Gaussian models to the region in the \emph{uv}-domain.

3C 84 is a highly complex source and adding more components to the map would have improved the reduced $\chi^{2}$ considerably. However, in this case, we are only interested in the relative change in the reduced $\chi^{2}$ value in order to estimate the degree to which the map prefers a single or double component fit to the region in question. For this reason, we only fitted the minimum number of components. In practice, this required a single Gaussian to be first be fit to the C1 region, as it dominates the map. Then we fit either one or two components to the region we were analyzing. We then compared the reduced $\chi^{2}$ values of the one and two component fits. Table \ref{modfit_compare} lists the reduced chi-square values. We found that in the cases of Split 1 and Split 2, the model preferred two components. However, in the case of Split 3, the algorithm favors  a single Gaussian. Adding a second Gaussian made only a marginal difference. Therefore, we consider Split 1 and 2 to be significant, but Split 3 to be marginal. In order to further explore the significance of the associations, we also tested the final model-fitted maps and tested the reduced $\chi^{2}$ values using a single or double component fit. The results were consistent with the previous method.

In order to determine the significance of these events being associated with $\gamma$-ray flaring, we follow the approached of \citet{rani2018}. From the simulated light curve analysis in Paper I, we found how often a $\gamma$-ray photon flux reached above the flux density, 2.4$\times10^{-7}$ph\,cm$^{-2}$\,s$^{-1}$, of the weakest flare from the de-trended 15-day binned $\gamma$-ray light-curve.
We found that for a given 15-day bin, there were $\sim$300 photons above this energy within the 5000 simulated light-curves. Given the monthly cadence of the VLBI observations, we therefore double this and estimate a 12\% chance of random association ($\sim$1.5$\sigma$). However, if we assume that the events are independent, the chance of random association for two ``split-flare-dissipate'' events is 1.4\% ($\sim$2.5$\sigma$) and 0.17\% ($\sim$3$\sigma$) for three ``split-flare-dissipate'' events.


\begin{table}
	\centering
	\caption{Comparison of Gaussian models. For each 'split-flare-dissipate' event, a single and double-component Gaussian model-fit was performed in order to determine if the map strongly prefers a single or double component structure. }
	\label{modfit_compare}
	\begin{tabular}{ccc} 
		\hline
		Split ID & Reduced $\chi^{2}$, 1-comp & Reduced $\chi^{2}$, 2-comp    \\
		\hline
         Split 1 & 211.7 & 140.2  \\
          Split 2 & 156.2 & 115.9  \\
          Split 3 & 96.2 & 86.2  \\
		\hline
	\end{tabular}
\end{table}

\begin{figure}
\includegraphics[width=0.99\linewidth]{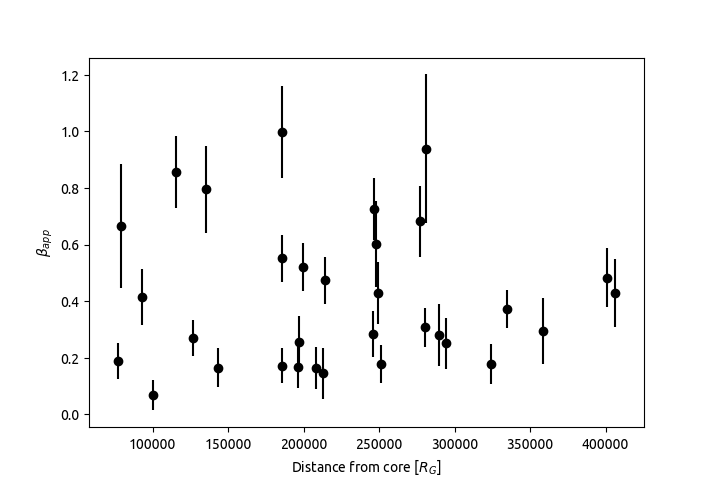}
\caption{Apparent jet speeds ($\beta_{\mathrm{app}}$) as a function of the projected distance in units of gravitational radii as measured from the WISE analysis. Note the lack of an accelerating trend.}  \label{accel}
\end{figure}

\subsection{Jet acceleration}\label{jetacc}

The estimated  apparent component speeds as a function of distance from the central nucleus are shown in Fig.~\ref{accel}. We find no evidence for significant acceleration within the central few thousand gravitational radii. Earlier studies at longer wavelengths have detected motions of $\sim$0.3-0.4\,mas/year ($\sim$0.3-0.5\,c), at distances much further downstream than in this study \citep{walker94,marr89,romney84,pt76}. This suggests that the jet has accelerated to its maximum speed within $<$125 000 $R_{\rm G}$ and stays at a roughly constant velocity for many further parsecs. 

\subsection{Jet viewing angle and Doppler factor}\label{view_angle}
The source shows evidence for apparent sub-luminal motions with the highest $\beta_{\rm app}$ found in Table \ref{kinematics} $\sim$0.9\,c.  Nevertheless, if the source is observed at the critical angle which maximizes the observed apparent speeds, we can estimate the minimum Lorentz factor $\Gamma_{\rm min} = \sqrt{1 +\beta_{\rm app}^{2}} \sim 1.35$. This alone cannot provide constraints on the viewing angle and Doppler factor of the source.\\

An alternative way to estimate the viewing angle is to use component EB2. If the emission from flare G4 is associated with the component EB2, then there is a time delay between the peaks of the $\gamma$-ray and radio emission. If this time delay is due to opacity effects, we can make the assumption that the component could be synchrotron self-absorbed and therefore derive another estimate of the viewing angle. Opacity effects due to SSA have been used to estimate the location of $\gamma$-ray emission relative to the radio emission in many sources \citep[e.g.][]{fuhrmann14}, assuming that the source is in equipartition:
\begin{equation}\label{coreshift}
    \Delta r = \frac{\beta_{\mathrm{app}}c \tau}{\sin \theta},
\end{equation}
where $\Delta r$ is the distance between the $\gamma$-ray and radio emitting regions, $\tau$ is the source frame time delay between the $\gamma$-ray and radio emission peaks and $\theta$ is the viewing angle to the source.  Component EB2, which we believe to be the origin of radio flare R3, peaks in October 2016. This leads us to estimate $\tau$ as $\sim$350 days. The component has moved $\sim$1\,mas $\sim$ 0.38\,pc in that time giving an estimate of $\Delta r$ and a speed of $\sim$0.65\,c. This leads to an estimate of $\sim$30$^{\circ}$ for the viewing angle, which is in tension with the measurement of $\sim65^{\circ}$ by \citep{fujita17} and below the previously derived lower limit. This could be evidence that the jet is not in equipartition.

%

\section{Discussion}

\subsection{Location of gamma-ray emission regions}
We presented multiple approaches to pinpoint the location of high-energy emission in 3C~84. The light curve analysis (presented in paper I) suggests the presence of multiple gamma-ray emitting sites in the source with the rapid variations being produced closer to the central engine (C1 region), while the long-term rising trend appears to be coming from the region further out in the jet (C3 region). Except from flare G4, which coincided with the radio flare, R3, in the C3 region, the light curve analysis did not conclusively determined the location of flares G1 to G3.
The jet kinematic analysis (this paper) also supports the presence of multiple gamma-ray locations.  Our analysis suggests that flare G1 is most likely associated with merging of components (E2 and W1) in the eastern lane of the jet. The component W1 is not detected furthermore after the merging, while component E2 continues to brighten and enlarge before it splits into two components and finally dissipates. The `split-flare-dissipate' behavior is observed three times and seems to be associated with flares G1, G2 and G3, respectively.  It is important to note that besides the most prominent $\gamma$-ray flares (G1 to G4), there are several low-amplitude flares observed before and after the major ones. It appears that both the `component merging' and the `split-flare-dissipation' seems to produce multiple $\gamma$-ray flares.

Other than being associated with the jet activity in the western lane,  the flare G4 (brightest among all $\gamma$-ray flares) appears to have a substantially different behavior to that of flares G1, G2 and G3. The VLBI kinematics has been recently studied by \citet{kino18}, where they reported that the jet ``flipped'' (where the unresolved brightest point in the jet rapidly changed location) during the flare G4/R3 period. Our analysis shows that the ``flip'' is in fact the emergence of the new component EB2 in a new location compared to the previous activity. 

In the following sections, we test several possibilities for the mechanism powering the $\gamma$-ray emission being in different active regions in the jet.

\subsection{Gamma-ray radiation mechanisms and acceleration processes}\label{mechanism}





There are several mechanisms that could potentially explain the observed behavior such as internal shock mechanisms \citep[e.g.][]{canto13} and magnetic reconnections \citep[e.g.][]{zhang11, mizuno11} including jet-in-jet formation (``mini-jets'') \citep[e.g.][]{giannios09}. Reconnections require reverse fields and can occur when a current sheet fragments into smaller plasmoids \citep{loureiro12}. These plasmoids can combine to form ``monster'' plasmoids, which can power large high-energy flares \citep{giannios13}. The reversed fields can cause a rapid rearrangement of the magnetic field topology which then can lead to efficient particle accelerations. The spine-sheath scenario has also been invoked by \citet{tavecchio14} to explain the high energy emission in 3C\,84, but this is considered unlikely by itself as it requires a smaller viewing angle to the source than is supported by observations. If the $\gamma$-ray emission is occurring closer to the SMBH, the emission could be produced in magnetospheric gaps \citep{aharonian17,magic_3c84_18}.


\begin{figure}
\includegraphics[width=0.99\linewidth]{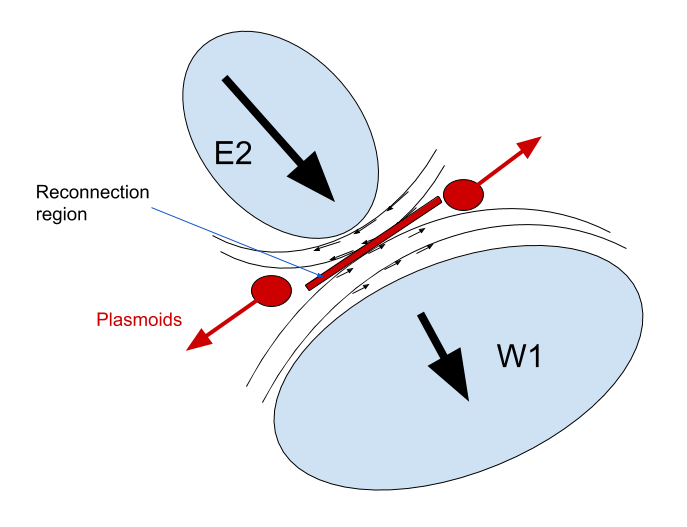}
\caption{A sketch of our interpretation of the behavior seen in 3C\,84.  Component W1 moves sub-relativistically, with turbulence behind the shock-front creating the slow-rising radio and $\gamma$-ray emission (as described in Section \ref{mechanism}). Component E2 moves mildly relativistically and collides with W1. With B-fields in different directions, the collision provides the inward force for large plasmoids to be formed (explaining the "split-flare-dissipate" behavior) that leads to the big flares G1,G2 and G3. Seed photons come from either the jet itself or the dense clumpy medium (or, more likely both).} \label{reconn_sketch}
\end{figure}

We begin by assuming that the C3 region is a traveling disturbance in the jet, originating from the jet launching region, near the central SMBH. Initially, after ejection, the magnetic field in the traveling shock would be weak because it is not yet strongly interacting with the external medium. Magnetic reconnections would therefore be rare. As the traveling shock interacts with the external medium, turbulence would be created behind the shock-front, which would then become progressively stronger \citep[as has been suggested in 3C\,84 by][]{nagai17}. The increased turbulence would locally amplify the magnetic field as filamentary structures and randomly orientated magnetic field strengths would lead to increased tangling which would therefore lead to increasing numbers of magnetic reconnection events. So long as fresh turbulence is introduced behind the shock front, the total magnetic field strength should increase and thus more reconnection events will occur. A strong magnetic reconnection event could cause a mini-jet (e.g. due to the collision of components E2 and W1) \citep{nalewajko11,giannios13}. In this scenario, when opposing polarity magnetic field lines meet, they would be annihilated. This would release large amounts of magnetic energy which heats the plasma and therefore accelerates particles, which in turn leads to $\gamma$-ray radiation \citep[e.g.][]{giannios13}. A further prediction is that as the underlying $\gamma$-ray flux increases, the strength of flares also increases, which is also seen (with further TeV detections now seen in 3C\,84).


Given the possibility that magnetic reconnection induced mini-jets may be important for producing the rapidly varying emission, a turbulence-induced magnetic reconnection scenario could also be a candidate for explaining the slow rising $\gamma$-ray trend \citep{mizuno11,
inoue11, zhang11, mizuno14}, where the C3 region is interpreted as a turbulent
shock front \citep{hiura18}.

The magnetic reconnection mini-jet combined with internal shocks and turbulence could provide a good potential scenario for much of the behavior seen in 3C\,84.  In Fig. \ref{reconn_sketch}, we sketch a scenario which could explain  ``split-flare-dissipate'' behavior seen in the cases of flares G1, G2 and G3. Component E2 collides with the slowly moving component W1, where the current sheets could be oriented north-south with reverse polarities along the interface of the shocks, causing the split due to reconnection to occur in the east-west direction. The flares G2 and G3 are then the same and the change in the direction of the split could be tracing the change in the orientation of current sheets. However, \citet{magic_3c84_18} analyzed the source and found that there was not enough jet power under the assumption of a magnetic field in equipartition for the ``jet-in-jet'' scenario and placed the emission closer to the SMBH. However, our analysis suggests the emission is downstream, which may indicate a magnetic field above the equipartition level when the jet was launched. There is also potential evidence for internal shock (e.g. the scenario described in Fig. \ref{reconn_sketch}) induced $\gamma$-ray behavior, which may imply that both mechanisms are producing $\gamma$-ray emission in both regions simultaneously. Further, \citet{giannios13} showed that ``mini-jets'' should occur within roughly 0.3-1 pc of the central engine, which is consistent with our observations, although the reconnection region itself is predicted to be less than 0.003\,pc, which is well below the resolution of our observations. In the next sub-sections, we compare our observational results with these predictions.

\begin{figure}
\includegraphics[width=0.99\linewidth]{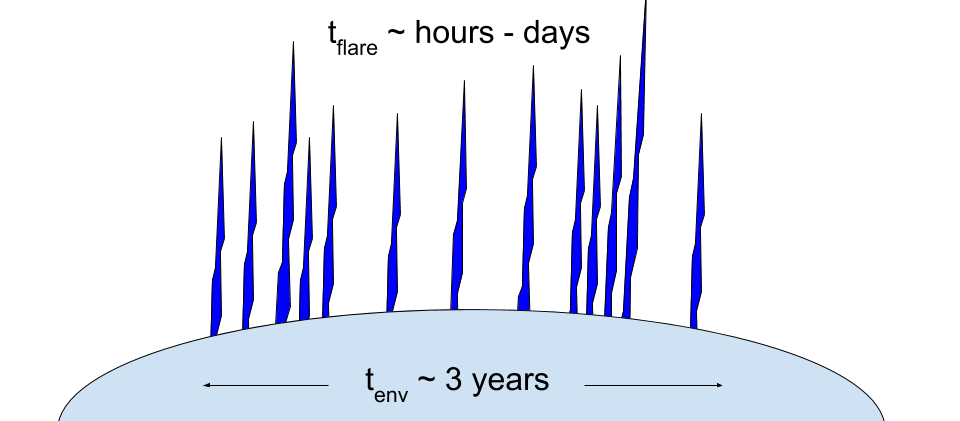}
\caption{A sketch of some of the observable quantaties for testing the magnetic reconnection scenario in 3C\,84 (see Section \ref{C3test} for more details). The envelope timescale $t_{\mathrm{env}}$ is considered roughly to be the slowly increasing emission seen from $\sim$2014-2017. $t_{\mathrm{flare}}$ is the timescale of the rapid variations seen on top of this.  } \label{tenv_sketch}
\end{figure}

\subsubsection{Testing in C3}\label{C3test}

The magnetic reconnection mini-jet scenario, where fast flaring above a slow envelope is predicted, has been analyzed from a theoretical perspective by \citet{giannios13}, presenting analytical expressions that can be tested in 3C\,84. In this paper, the author assumed that the slow envelope emission would be on the order of days, however we are equating the slow-rising trend with the envelope emission with an envelope timescale of $\sim$years (e.g. see Fig. \ref{tenv_sketch}). We have one key observable, the dissipation radius $R_{\mathrm{diss}}$ where most of the high energy emission is emitted. This can be directly derived from the data. In the case of 3C\,84, this emission could be occurring within the C3 region when it was at $\sim$2-3\,mas from the core, equivalent to a projected distance $\sim$1\,pc. The difference between the projected and de-projected distances should not be very different if the viewing angle is relatively large (as is suggested in Section \ref{view_angle}). However, if the viewing angle is smaller, the projected and de-projected distances will vary by a factor of $\sin \theta$. As discussed in the previous section, since 3C 84 is only mildly relativistic, we assume that the Lorentz factor and Doppler factors are $\sim$1, although a mildly higher Doppler factor should not affect our conclusions significantly. Primed quantities are in the observer frame. The mildly relativistic bulk jet emission may have mini-jets that have their own Doppler factor $\delta_{j}$. The timescale $t_{\mathrm{env}}$ of the envelope in the jet frame is:
\begin{equation}
    t_{\mathrm{env}}' = \frac{(1+z) R_{\mathrm{diss}}}{\Gamma^{2} \beta},
\end{equation}
where $\delta$ is the relativistic Doppler factor of the jet. We introduce a factor of $\beta$ here as in the paper of \citet{giannios13} which assumed it to be one. Using cgs units, we set $R_{\mathrm{diss}} \sim 1 \mathrm{pc} \sim 3\times10^{18}$\,cm, giving an envelope timescale of $t_{\mathrm{env}} \sim 1\times10^{8} \mathrm{s} \sim 3$\,years. This matches well with the observed envelope in 3C\,84, which began in approximately 2013 and continues until 2015 (see Fig. \ref{G_lc} and the sketch in Fig. \ref{tenv_sketch}). According to this model, when the reconnection timescale, which is directly related to the Schwarzchild radius of the SMBH ($R_{\mathrm{sch}}$), becomes comparable to the expansion timescale of the jet, much of the high-energy emission is dissipated (i.e. at $R_{\mathrm{diss}}$). This is because the model assumes that small scale magnetic field structures are embedded into the jet on the scale of of the black-hole event horizon. It then follows that we can then solve for the magnetic reconnection speed $\epsilon$ in units of c:
\begin{equation}
    \epsilon = \frac{\Gamma^{2}R_{\mathrm{sch}}}{R_{\mathrm{diss}}}.
\end{equation}
Using $M_{\mathrm{BH}} \sim 3\times10^{8} M_{\odot}$, leading to $R_{\mathrm{sch}} \sim 1\times10^{14}$\,cm, this gives $\epsilon \sim 3 \times 10^{-5}$. This value is several orders of magnitude lower than the expected value of $\sim$0.1 used in \citet{giannios13}. Nevertheless, we can use this value with the envelope timescale to estimate the size of the of the reconnection region in the jet frame, $l'$:
\begin{equation}\label{l_dash}
    l' = t_{\mathrm{env}} \Gamma \epsilon c,
\end{equation}
and therefore the jet frame flare timescale $t_{\mathrm{flare}}$:
\begin{equation}\label{t_flare}
    t_{\mathrm{flare}}' = \frac{0.1 l'}{\delta_{p} c},
\end{equation}
where $\delta_{p}$ is the Doppler factor of a mini-jet plasmoid. This leads to an jet frame reconnection region size of $l' \sim 9\times10^{14} \mathrm{cm} \sim 3\times10^{-4}$\,pc and a flare timescale of $t_{\mathrm{flare}} \sim 60$\,mins. Since relativistic effects are only mild in 3C\,84, the observer and jet frame emission will be quite similar. Even if we assume that the Doppler factor of a plasmoid is also of $\sim$1, the predicted flare timescale is somewhat shorter than the observed timescale of $\sim$hours to days. However, it is likely that mini-jets oriented towards the line-of-sight of the observer would have Doppler factors greater than unity.\\

While the picture fits in a broad sense, there are also large inconsistencies. For example, the size of the reconnection region, which should be related to the SMBH size by $l'\simeq\Gamma R_{\mathrm{sch}}$, and is an order of magnitude larger than the size estimated via the location of the dissipation region. Similarly, the estimated flare timescale is roughly an order of magnitude faster than what is seen. The reconnection rate is two orders of magnitude lower than expected. In our calculations, we assumed a SMBH mass of $\sim3 \times 10^{7} M_{\odot}$ \citep{onori17}. If the mass of the SMBH were higher than this, it could resolve the discrepancy. For example, a SMBH mass of $\sim10^{10} M_{\odot}$ would reconcile the values. However, it is possible that the mass is lower, of order $10^{6}-10^{7} M_{\odot}$ in 3C\,84, which would make the inconsistencies worse, \citep[see][for a discussion about the SMBH mass in 3C\,84]{sani18}. Another possibility is that other physics is involved such as radiation \citep[e.g.][]{uzdensky15}. The presence of a spine-sheath structure could also potentially create the conditions for magnetic reconnections if the current sheets are opposed in the regions where the spine and sheath meet. \\

We consider it more likely that the reconnection length scale would not scale directly with $R_{\mathrm{sch}}$. The turbulence scenario described at the start of Section \ref{mechanism} would naturally explain such a discrepancy, as randomly aligned ``cells'' of plasma could easily have opposing magnetic field directions \citep{marscher14}. We can then estimate the number of ``cells'' within a volume \citep{burn66}:
\begin{equation}
    n_{\mathrm{cells}} \sim \left(\frac{70\%}{P\%}\right)^2,
\end{equation}
where $P\%$ is the polarization degree of the emitting region. The size of a ``cell'', $R_{\mathrm{cell}}$, which should be equivalent to the reconnection size $l'$, would then simply be the third root of the volume of the region divided by the number of cells:
\begin{equation}
    R_{\mathrm{cell}} \sim \frac{R_{\mathrm{region}}}{n_{\mathrm{cells}}^{1/3}},
\end{equation}
where $R_{\mathrm{region}}$ is the size of the region. This assumes that the emitting region is roughly spherical. If the emitting region is disk-like, the area of the region should be divided by the number of cells. Thus, for a polarization degree of $\sim$2\%, which was localized within the C3 region \citep{nagai17}, this leads to $n_{\mathrm{cells}} \sim 1000$ and to a cell size (equivalent to a reconnection size) of $\sim$0.02\,pc $\sim6 \times 10^{16}$\,cm. Setting $l'=R_{\mathrm{cell}}$ in equations \ref{l_dash} and \ref{t_flare} allows us to estimate $\epsilon \sim 0.02$, which is closer to the expected value of ~0.1 and $t_{\mathrm{flare}} \sim 2$\,days, which better matches the observations.  Including a Doppler factor for the mini-jets ($\delta_{j}$) would further match the model with the observed behavior. \\

\subsubsection{Testing in C1}

Alternatively, we  consider the case of the high energy emission originating from the C1 region. In this case, we set $R_{\mathrm{diss}}$ to be the beam-size of $\sim$0.15\,mas. This leads to $R_{\mathrm{diss}} \sim 0.05 \mathrm{pc} \sim 1.5\times10^{17}$\,cm, which must be emphasized is an upper limit.
Using this limit, we find the envelope timescale to be $t_{\mathrm{env}} \sim 8$\,weeks and the reconnection speed to be $\epsilon \sim 7 \times 10^{-4}$. We then estimate the size of the reconnection region to be $l' \sim 1\times10^{14} \mathrm{cm} \sim 3.3\times10^{-5}$\,pc and a flare timescale of $t_{\mathrm{flare}} \sim 5$\,mins, which is inconsistent with the observations (fastest flare observed from the source has a timescale of $\sim$10~hr).   Faint polarized emission of $\sim 0.5\%$ was found by \citet{nagai17}, which would lead to $n_{\mathrm{cells}} \sim 20 000$ and a cell size of $\sim 2.4 \times 10^{-6}$\,pc. Using equation \ref{t_flare} we find $t_{\mathrm{flare}} \sim 30$\,seconds. Thus we find that if the emission is originating from within the C1 region, it is highly unlikely to be due to magnetic reconnections.

We can also use the observed jet and $\gamma$-ray luminosity to further constrain possible models. Using a peak $\gamma$-ray flux density of $\sim8 \times 10^{-7}$ph/cm$^{2}$/s, from the period before flare G4, we estimate the isotropic $\gamma$-ray luminosity to be $\sim9 \times 10^{37}$\,W (9 $\times 10^{44}$\,ergs/s). This is less luminous than the flaring period that was also analyzed by \citet{baghmanyan17}. \citet{aharonian17} set lower limits for the jet luminosity in the mini-jets scenario with \citet{magic_3c84_18} finding that this lower limit is $\sim4 \times 10^{37}$\,W (4$\times 10^{44}$\,ergs/s), using a more luminous flare. The less luminous flaring analyzed in this paper should further lower this limit. This would make the mini-jets scenario compatible with the jet luminosity estimated to be of the order $\sim0.3-5 \times 10^{37}$\,W (0.3-5 $\times 10^{44}$\,ergs/s), found by \citet{abdo09} and \citet{dunn_fabian04}.

\subsubsection{Source of the seed photons}

It is presumed that the specific mechanism to produce the $\gamma$-rays is via the inverse Compton mechanism in leptonic scenario, however hadronic process is equally probable. It is possible that the mini-jets themselves would have their own Doppler factor ($\delta_j$) \citep{mar14}, although it is not necessary to describe the observed behavior. The acceleration of high energy electrons could occur in very small regions or these small regions experienced short Doppler boosts (i.e. mini-jets, see Section \ref{C3test} and \citet{giannios13}). Both of these conditions can occur in magnetic reconnections arising in turbulent regions. In addition to such a Doppler factor, in order to produce the $\gamma$-rays, seed photons are required.

The two main sources for the seed photons are thought to be either the jet itself (synchrotron self-Compton, SSC) or the external environment (external Compton,  EC) \citep[e.g.][]{sokolov05}. The nature of the circumnuclear gas, which could provide the seed photons near the core of 3C\,84 has also been actively studied, with much of this gas in-flowing to the central kpc \citep{salome06,lim2008}. In the region closest to the SMBH, it is thought that free-free absorption suppresses counter-jet emission in VLBI images \citep{dhawan98,walker2000}. \citet{nagai17} has suggested from rotation measure measurements that this emission, if it is related to the accretion flow, must be highly inhomogeneous, consistent with an analysis of the free-free absorption opacity, suggested by \citet{fujita16}. Many previous studies have found evidence for a dense ionized plasma in the central regions of 3C\,84 \citep[][]{odea84,walker2000}, with the C3 region itself likely to be within the broad line region (BLR). \citet{punsly18} recently found evidence for broad H$\beta$ emission lines within the system. \citet{krabbe2000} investigated the near-infrared emission and found ionized gas, molecular gas and hot dust in the central regions of 3C\,84. These sources could all potentially provide seed photons that are external to the jet. Seed photons from the jet do the job equally well \citep{kino17}.

\subsection{Comparison with M 87}

M 87 is likely the best source to compare 3C\,84 with, as it is nearby and has also been analyzed using WISE \citep{mertlovm87,park19}. However, it is important to note that M87 is relatively closer, hence the observations provide a higher linear resolution. Additionally, the data used by \citet{mertlovm87} was of higher (every $\sim$21 days) cadence than these observations. Nevertheless, some interesting conclusions can be drawn. In their study, they found that the M\,87 jet is highly stratified, with potential evidence for a spine-sheath structure.
We observe two ``lanes'' of mildly-relativistic emission, which could be interpreted as a sheath, however we observe no motion within these ``lanes'' (see Fig.~3). Motion of faint features within the lanes would likely need to be detected in order to provide evidence for a spine. Currently the data does not show this.

Thus, we find potential evidence for a sheath but not a spine. Given the large viewing angle of the source, a spine might present but is de-beamed and thus too faint to detect. Furthermore, 3C 84 has a highly cylindrical jet beyond the C1 region. Given that \citet{fujita16} suggested that the density of the ambient medium could be quite low, the highly cylindrical outflow could suggest a strong magnetic sheath in 3C\,84 \citep[as suggested by][]{blandford18}.

Another point of comparison is the lack of acceleration seen in 3C\,84 (see Section \ref{jetacc}). Significant jet acceleration was observed in M87 by \citet{park19} and others. A possible way to reconcile the results is that the C1 region is not the jet launching region, but a recollimation shock at the Bondi radius.  A similarity, however,  is that we also have found a large scatter in the $\beta_{\mathrm{app}}$, which is similar to what is seen by \citet{park19}. The scatter could be due to the jet emission having multiple streamlines with different acceleration profiles. In the absence of acceleration, it could be interpreted as streamlines having multiple jet velocity profiles.

Finally, we notice a curvature in the west to south-eastern motions detected in the inner $\sim$mas of 3C\,84 (Fig. \ref{016mas4}). We suggest that this could be due to rotation of the jet base. This will be analyzed further in an upcoming paper based on Global mm-VLBI Array observations.

\section{Conclusions}\label{conclusions}

We presented a thorough analysis of the jet kinematics in 3C~84 over the period 2010 until 2017. The observed changes in the jet morphology were  compared with $\gamma$-ray flaring activity detected in the source in order to better understand the high-energy emission of the source.


     We find that the jet appears to have accelerated to its maximum mildly relativistic speeds within a $\sim125 000$\, gravitational radii of the jet launching point.  Subluminal components motions have been detected in both eastern and western lanes of the jet. 
    There is a large variation in the observed speeds for a given separation from the jet base, in both lanes of the jet. We found a maximum speed in the jet of $\sim$0.9\,c leading to a minimum Lorentz factor of $\sim$1.35.

     We noticed that when faster moving regions interact with slower moving regions, $\gamma$-ray flares are observed.
     Two hotspots are detected in the C3 region that become bright and then dissipate to the west.
     The second hotspot began to brighten in late 2015, and appears to be associated with a particularly large $\gamma$-ray flare (G4). Our study indicates that $\gamma$-rays are produced in both eastern and western lanes in the jet.
     We discussed the possibility of $\gamma$-ray flares being produced via  magnetic reconnection induced mini-jets and turbulence. Moreover, we find the evidence that there could be an excess of magnetic energy or gradients in pressure and the ambient medium.



\section*{Acknowledgements}
I would like to thank Jae-Young Kim and Thomas Krichbaum for their valuable discussions and contributions to this paper.
This work by Jeffrey A. Hodgson was supported by Korea Research Fellowship Program through the National Research Foundation of Korea(NRF) funded by the Ministry of Science and ICT(2018H1D3A1A02032824). He is also supported via the National Research Foudnation of Korea (NRF) grant (2021R1C1C1009973). Y.M. acknowledge support from the ERC synergy grant ``BlackHoleCam:
Imaging the Event Horizon of Black Holes'' (Grant No. 610058). S.-S. Lee was supported by a National Research Foundation (NRF) of Korea grant funded by the Korean government (MSIT No. NRF-2020R1A2C2009003).  This study makes use of 43 GHz VLBA data from the VLBA-BU Blazar Monitoring Program (VLBA-BU-BLAZAR; http://www.bu.edu/blazars/VLBAproject.html), funded by NASA through Fermi Guest Investigator grant 80NSSC17K0649. The VLBA is an instrument of the National Radio Astronomy Observatory. The National Radio Astronomy Observatory is a facility of the National Science Foundation operated by Associated Universities, Inc.
J.P. acknowledges financial support from the Korean National Research Foundation (NRF) via Global PhD Fellowship Grant 2014H1A2A1018695. J.P. is supported by an EACOA Fellowship awarded by the East Asia Core Observatories Association, which consists of the Academia Sinica Institute of Astronomy and Astrophysics, the National Astronomical Observatory of Japan, the Center for Astronomical Mega-Science, the Chinese Academy of Sciences, and the Korea Astronomy and Space Science Institute.




\bibliographystyle{mnras}
\bibliography{Bibliography} 










\clearpage

\appendix

\section{Additional plots and tables from WISE analysis}

\begin{figure*}
\includegraphics[width=0.47\textwidth]{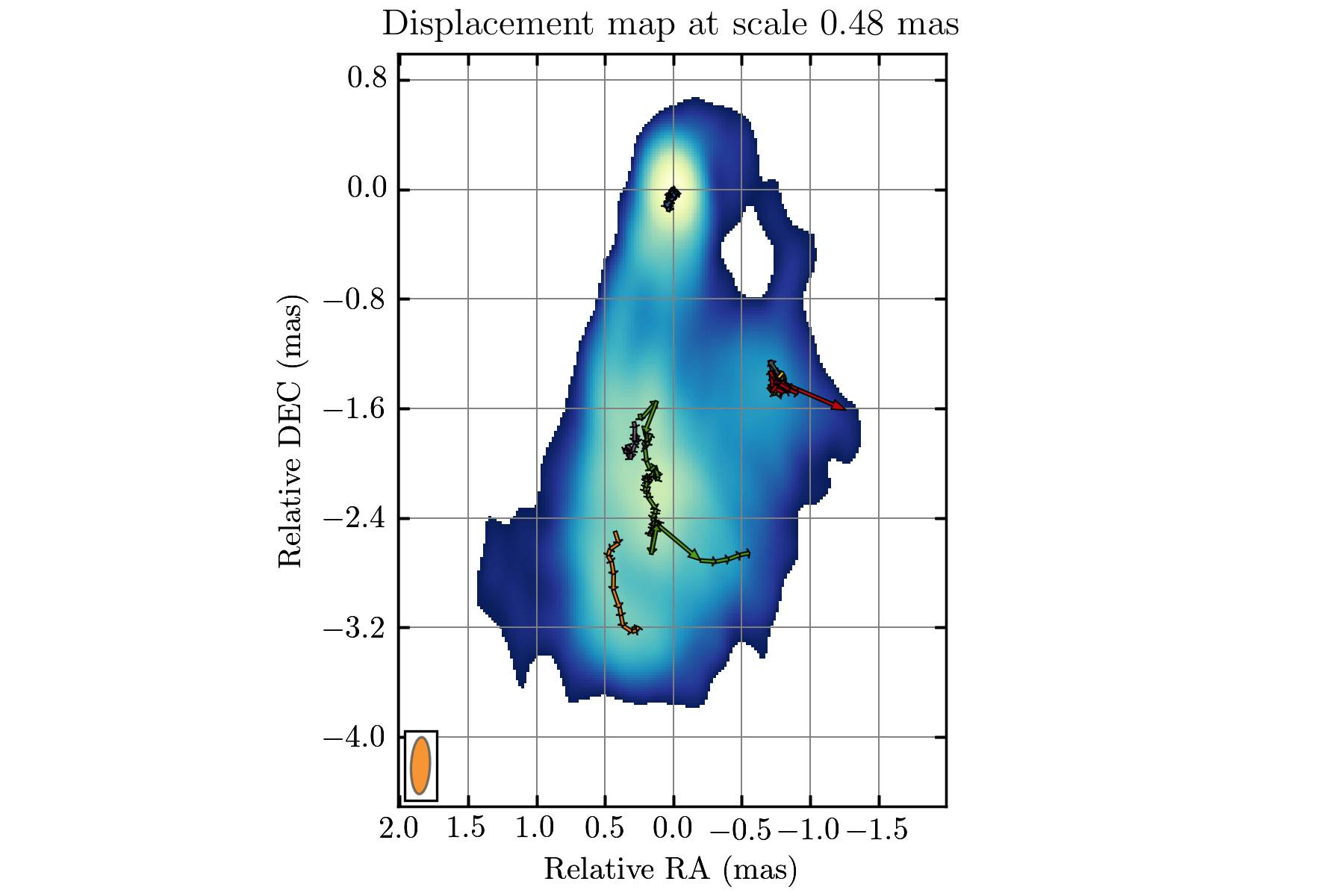}
\includegraphics[width=0.47\textwidth]{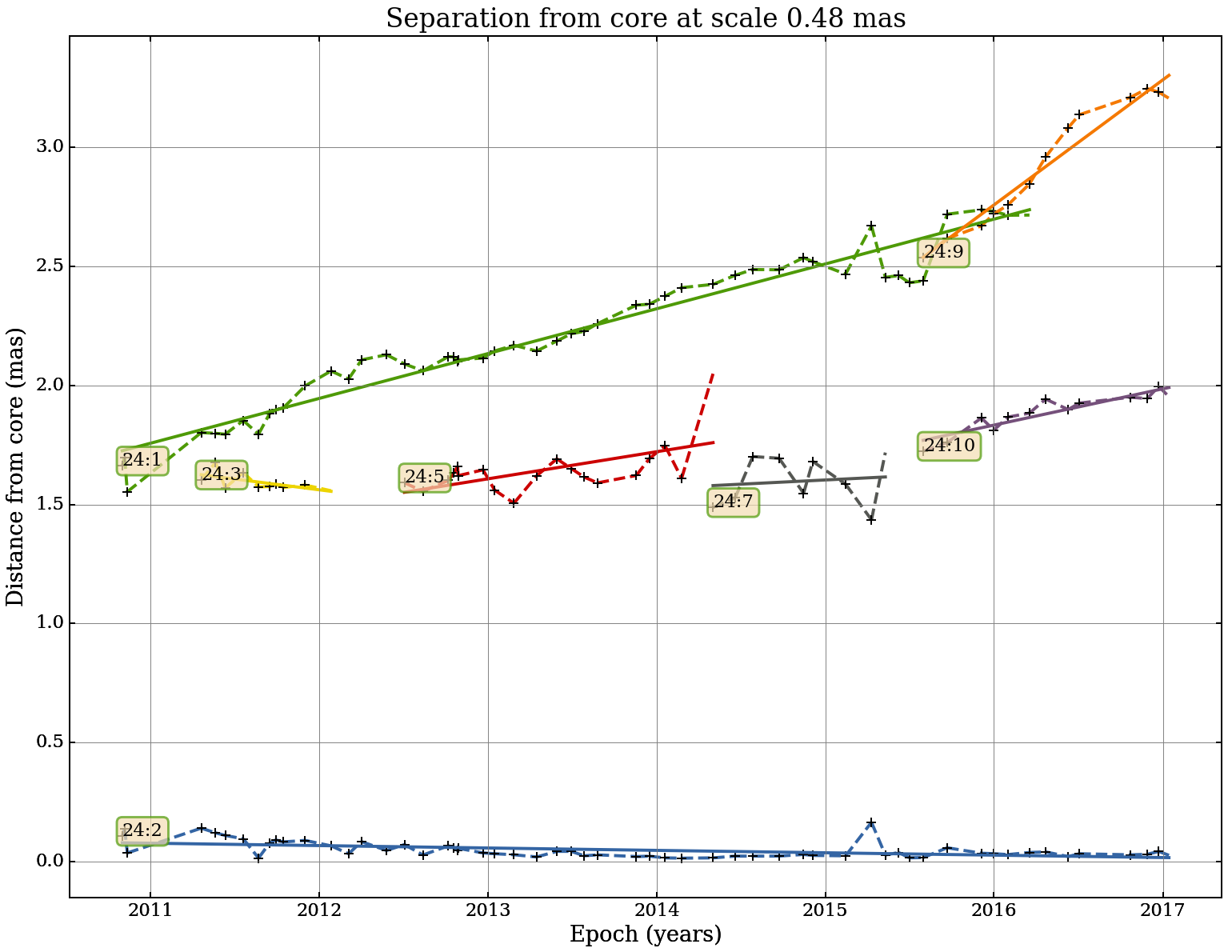}
\caption{\textbf{Projected kinematics (left) and fitted core-separation plots (right) with length scale 0.48\,mas, 8 component fit. } }\label{048mas8}
\end{figure*}

\begin{figure*}
\includegraphics[width=0.47\textwidth]{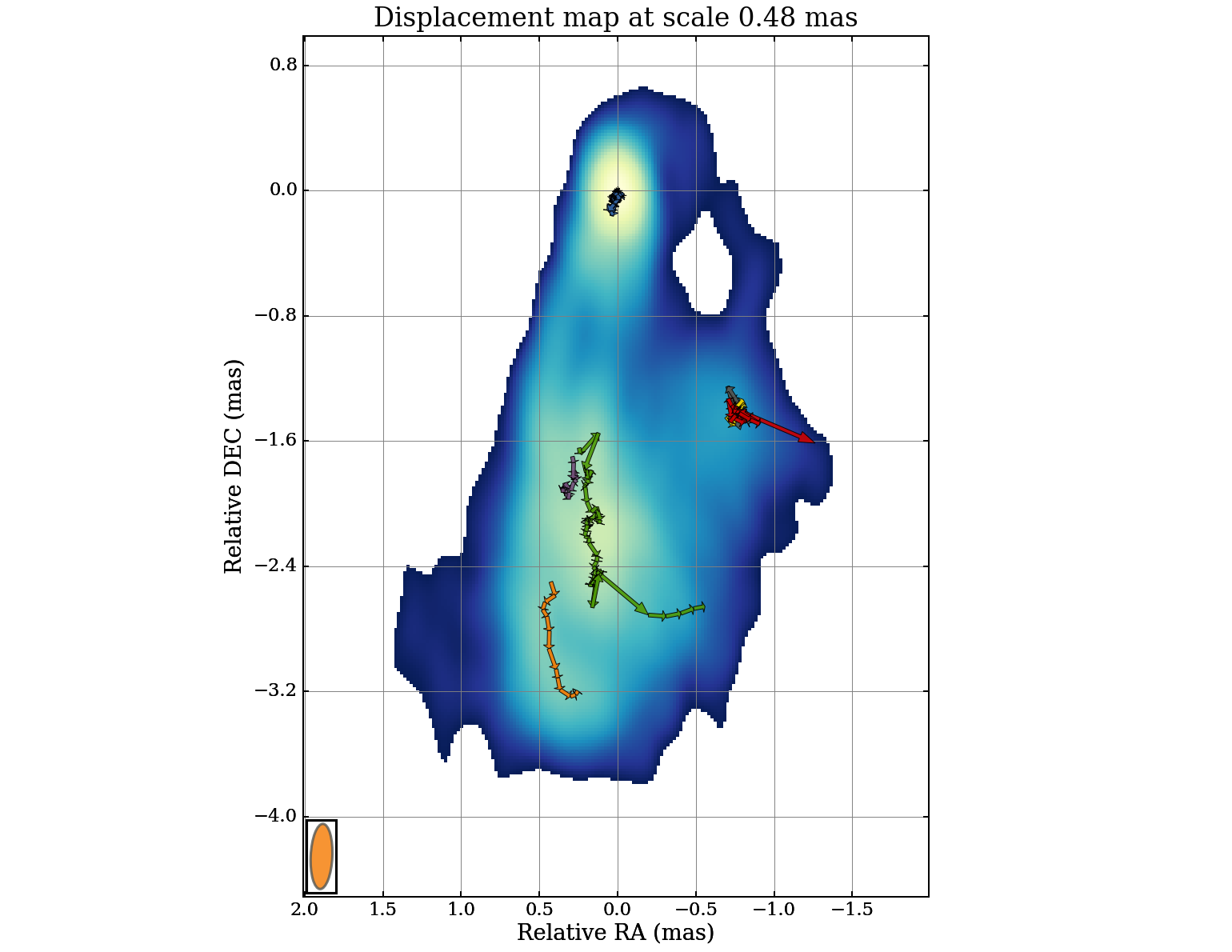}
\includegraphics[width=0.47\textwidth]{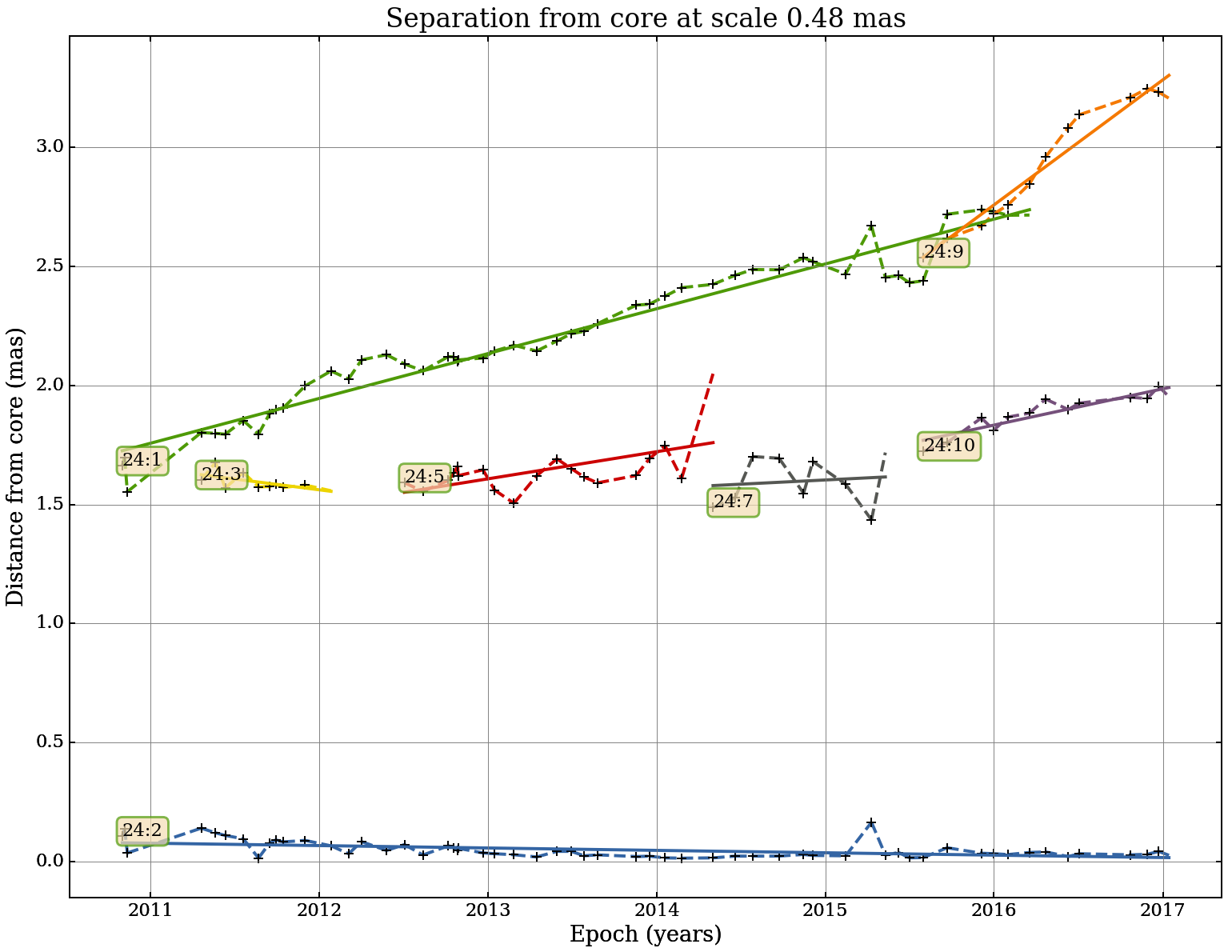}
\caption{\textbf{Projected kinematics (left) and fitted core-separation plots (right) with length scale 0.48\,mas, 4 component fit. } }\label{048mas4}
\end{figure*}

\begin{figure*}
\includegraphics[width=0.47\textwidth]{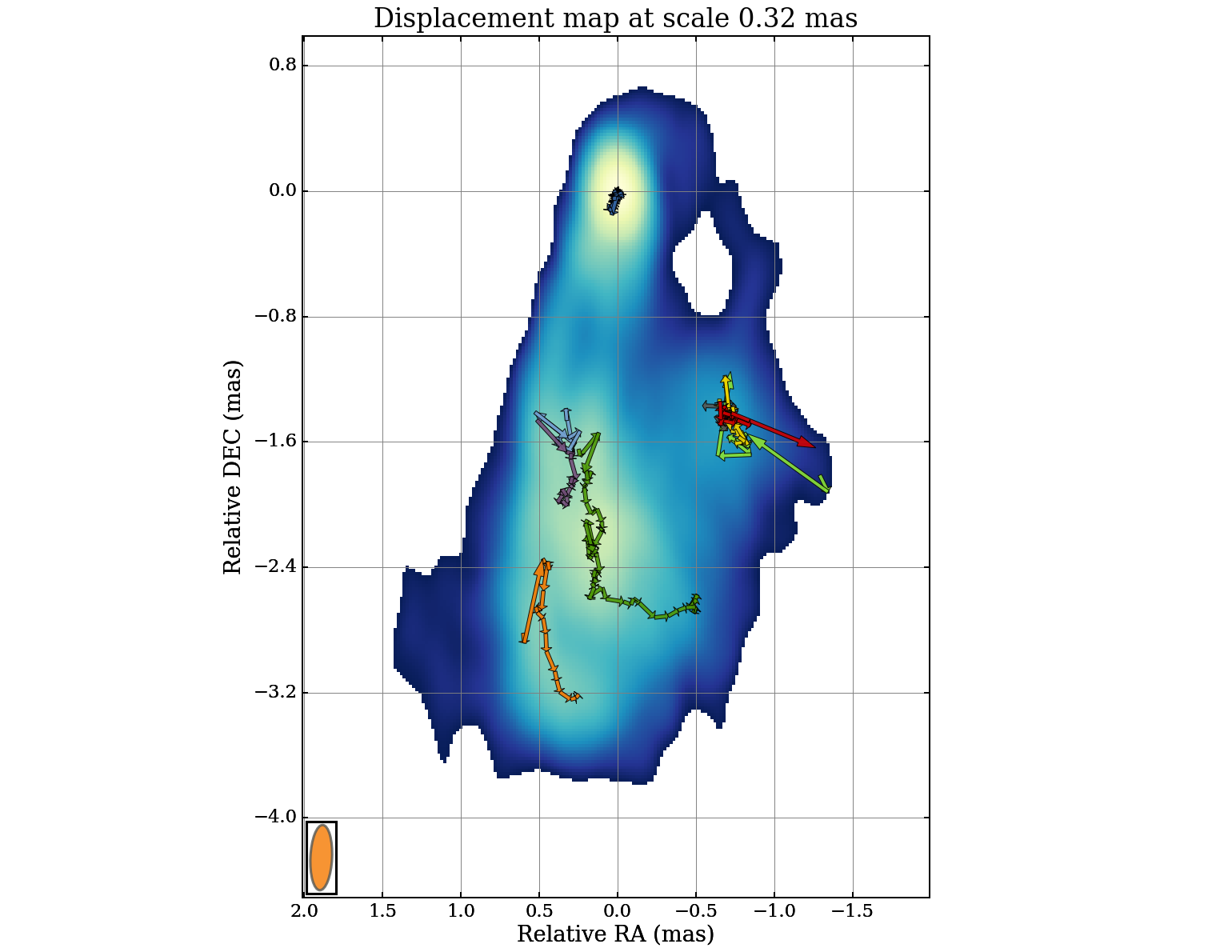}
\includegraphics[width=0.47\textwidth]{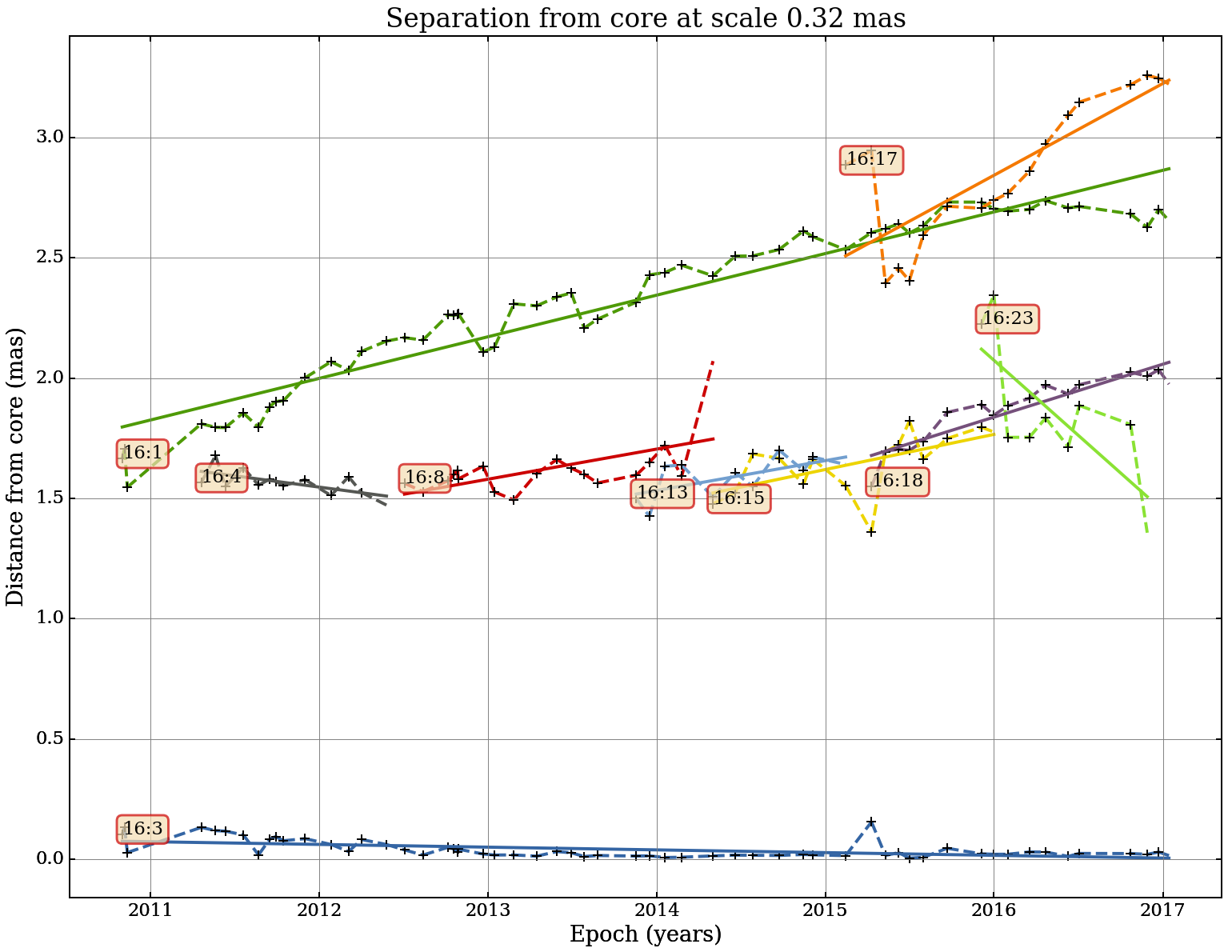}
\caption{\textbf{Projected kinematics (left) and fitted core-separation plots (right) with length scale 0.32\,mas, 8 component fit. } }\label{032mas8}
\end{figure*}

\begin{figure*}
\includegraphics[width=0.47\textwidth]{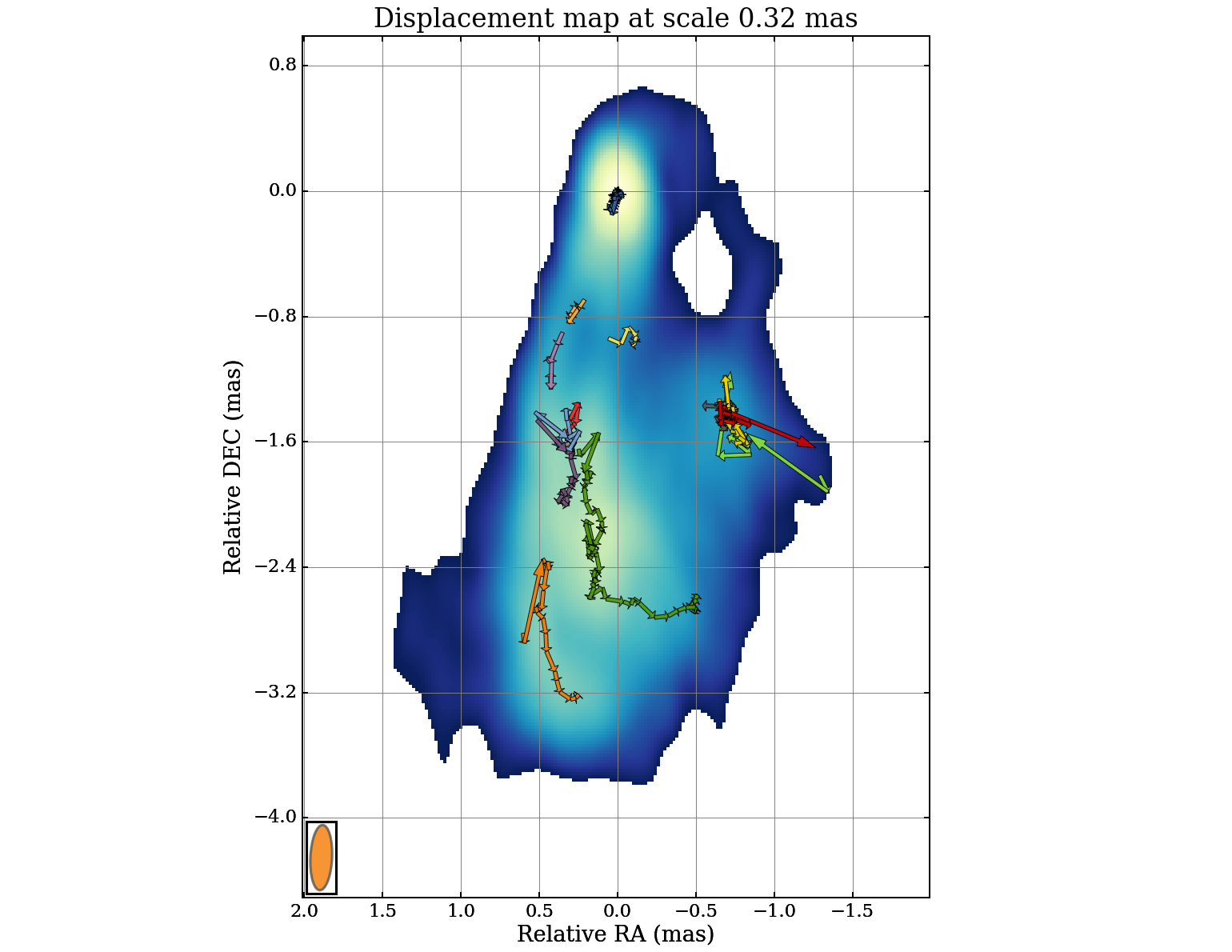}
\includegraphics[width=0.47\textwidth]{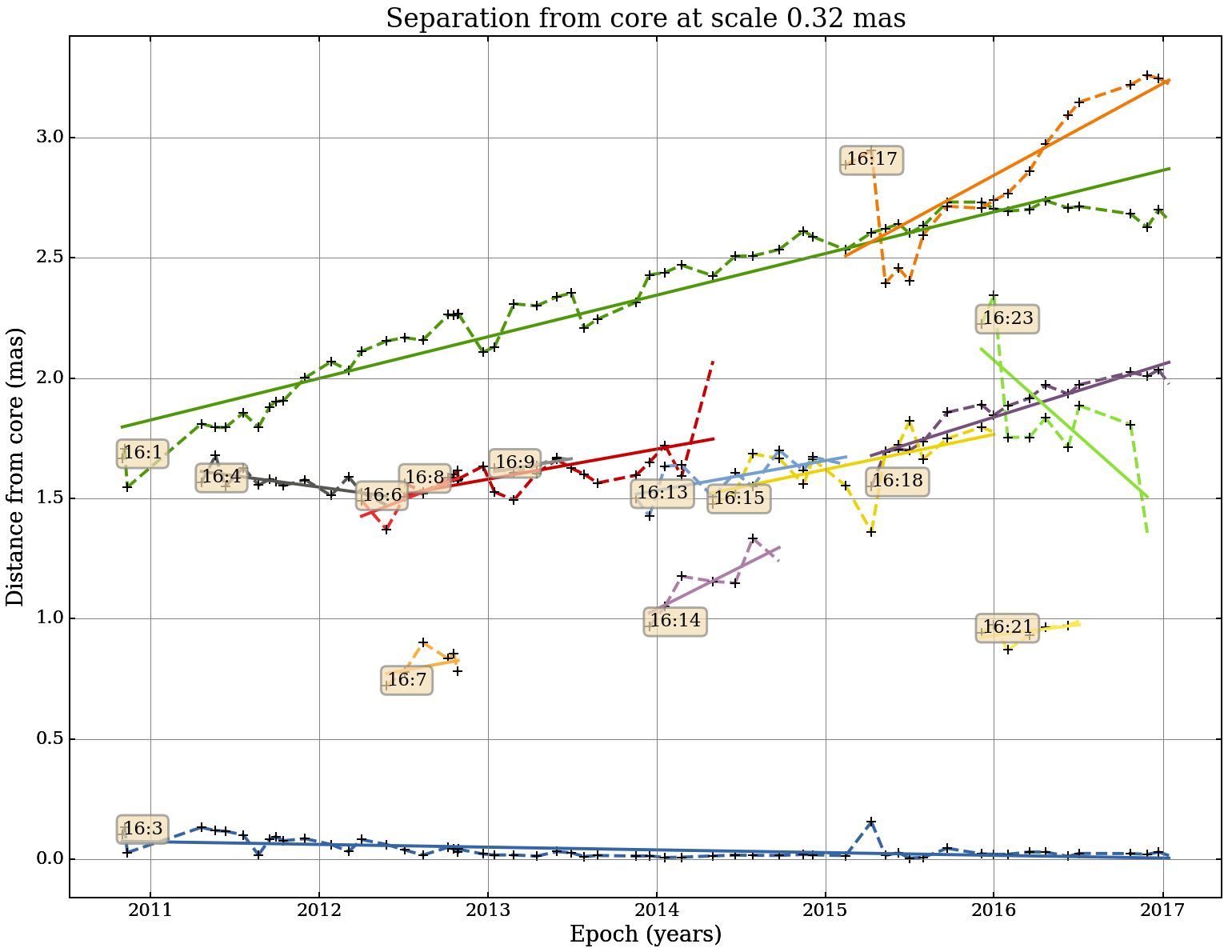}
\caption{\textbf{Projected kinematics (left) and fitted core-separation plots (right) with length scale 0.32\,mas, 4 component fit. } }\label{032mas4}
\end{figure*}

\begin{figure*}
\includegraphics[width=0.47\textwidth]{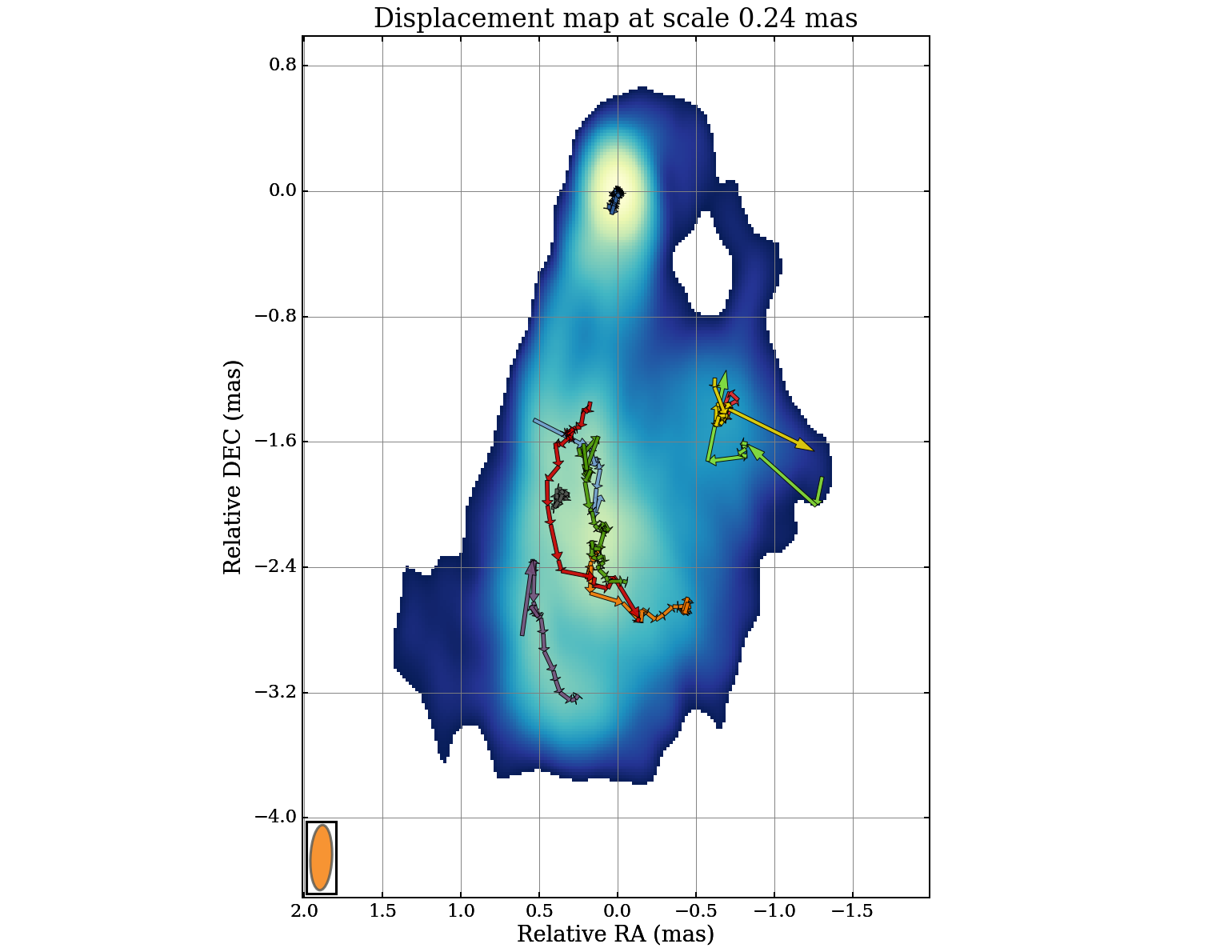}
\includegraphics[width=0.47\textwidth]{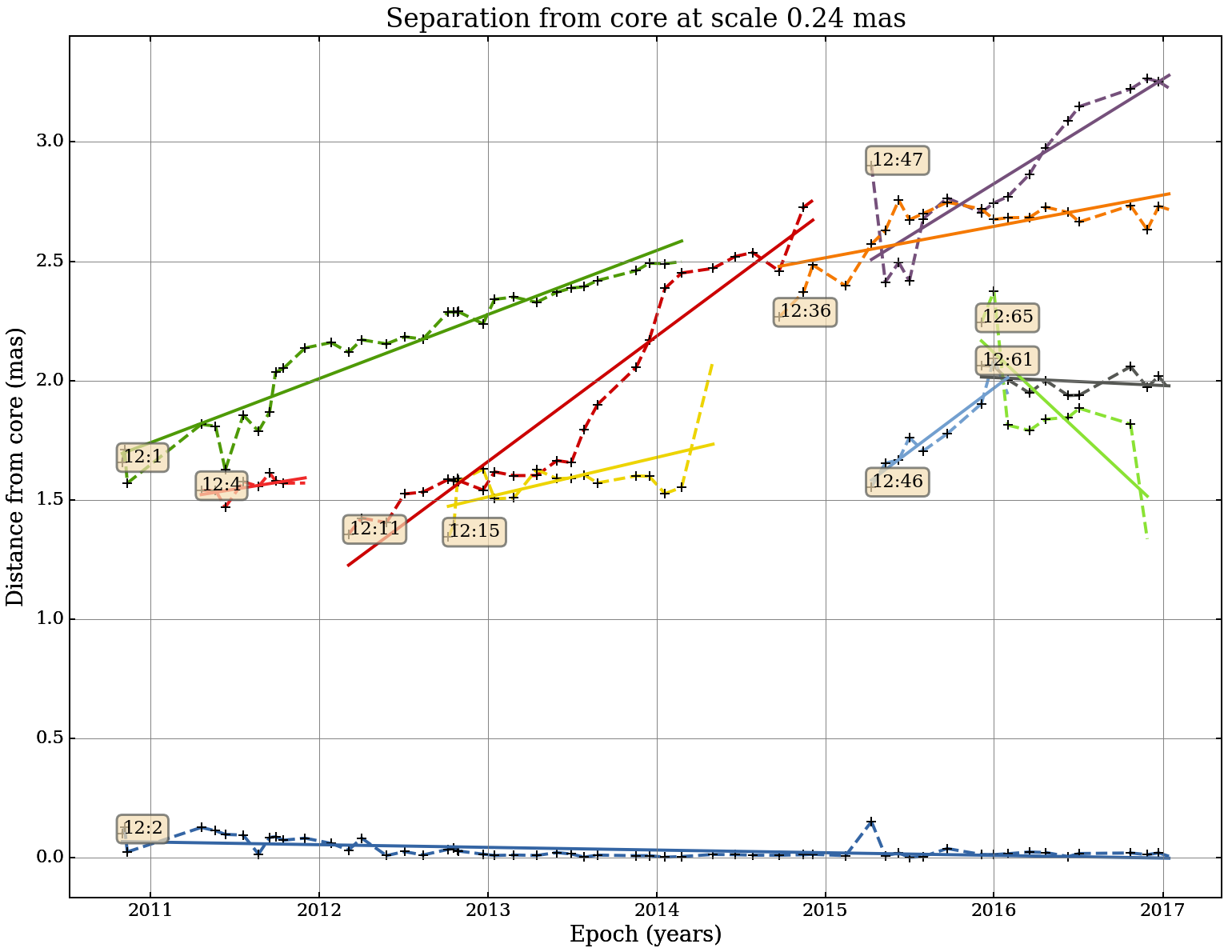}
\caption{\textbf{Projected kinematics (left) and fitted core-separation plots (right) with length scale 0.24\,mas, 8 component fit. } }\label{024mas8}
\end{figure*}

\begin{figure*}
\includegraphics[width=0.47\textwidth]{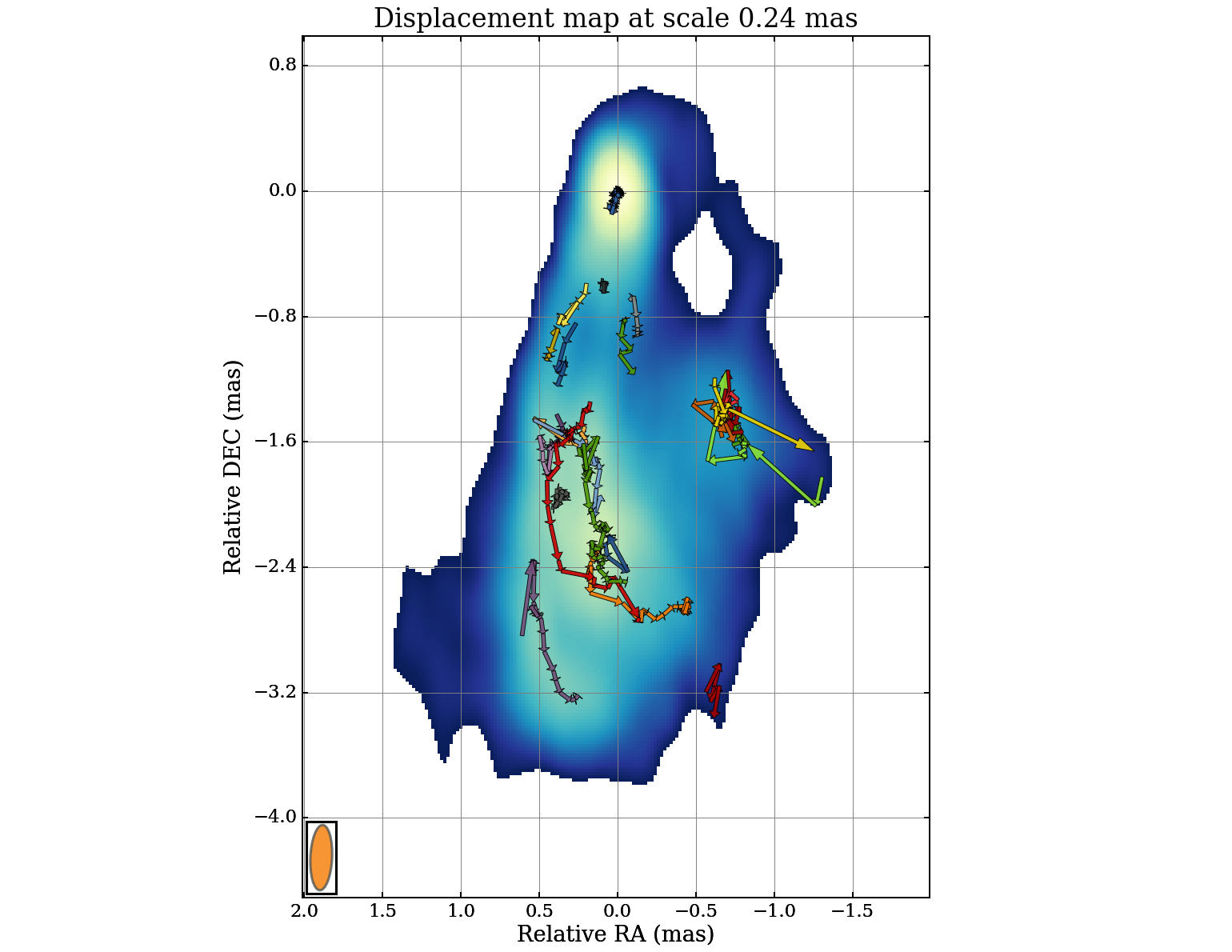}
\includegraphics[width=0.47\textwidth]{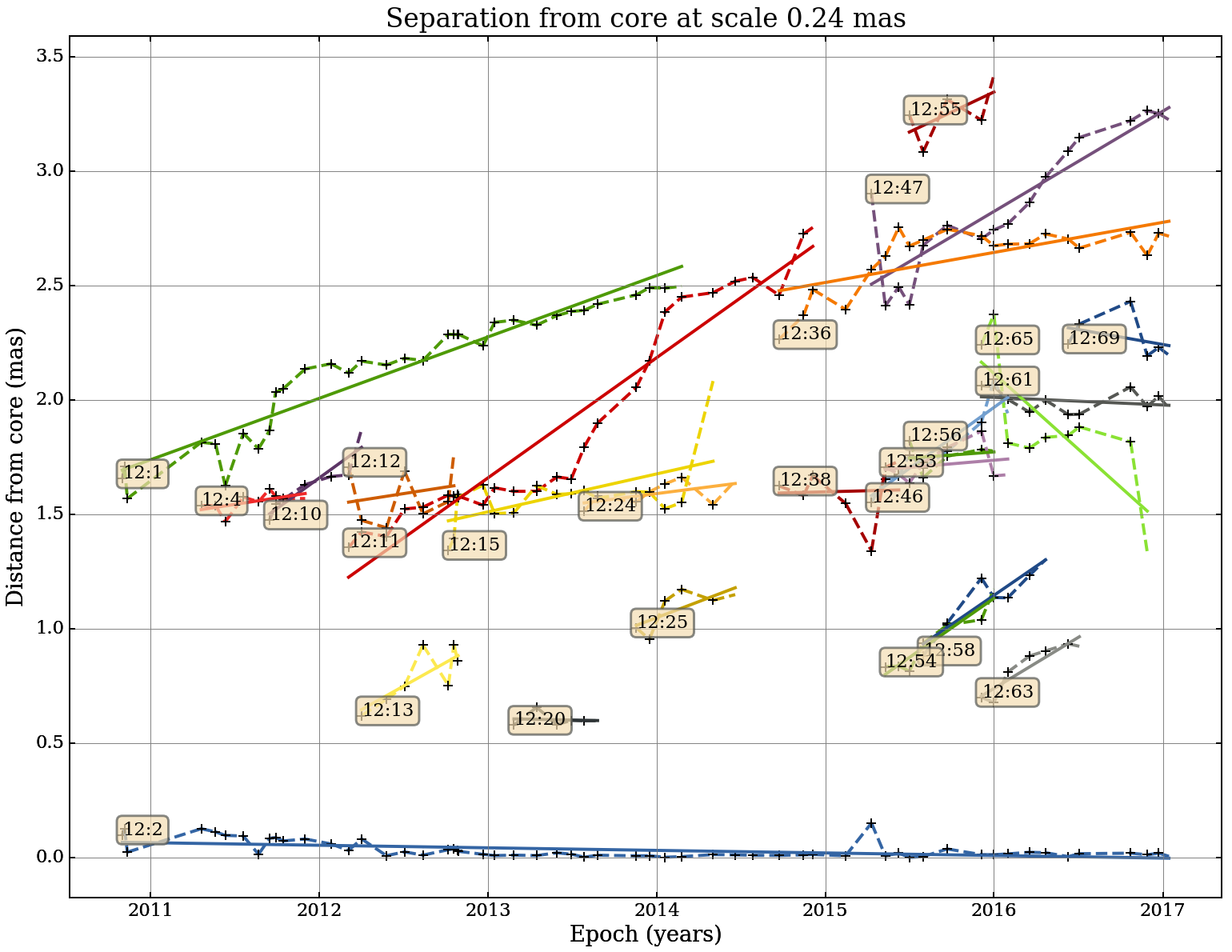}
\caption{\textbf{Projected kinematics (left) and fitted core-separation plots (right) with length scale 0.24\,mas, 4 component fit. } }\label{024mas4}
\end{figure*}

\begin{figure*}
\includegraphics[width=0.47\textwidth]{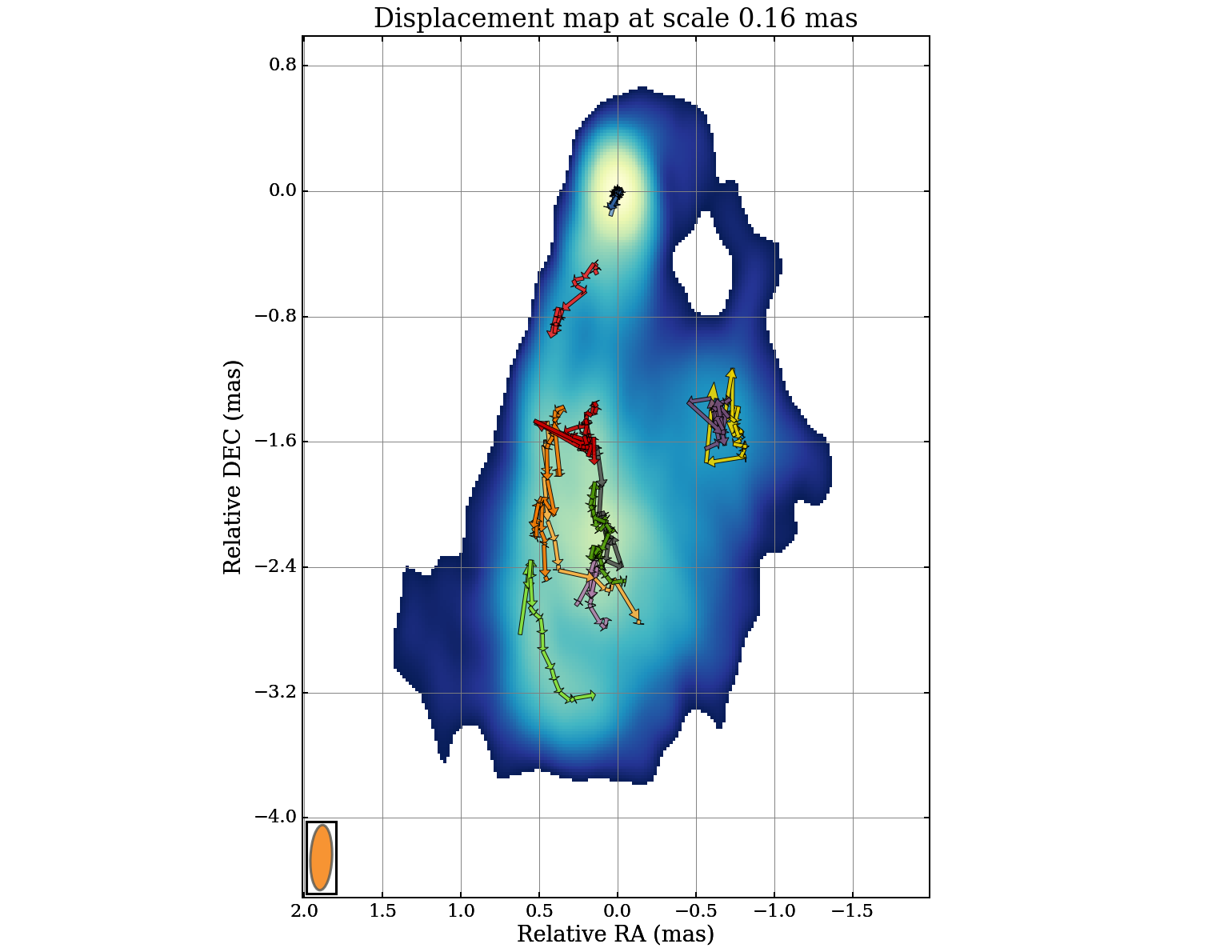}
\includegraphics[width=0.47\textwidth]{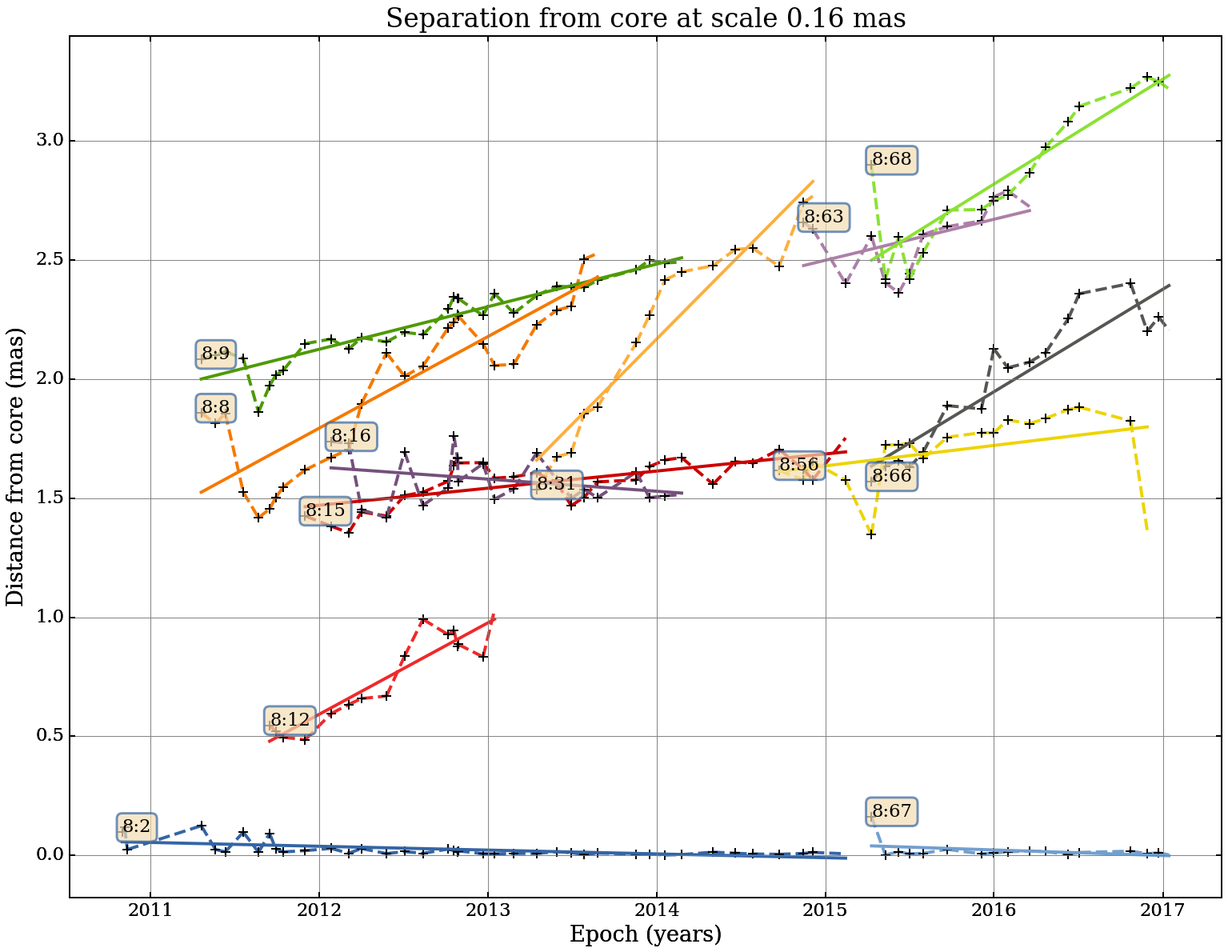}
\caption{\textbf{Projected kinematics (left) and fitted core-separation plots (right) with length scale 0.16\,mas, 12 component fit. } }\label{016mas12}
\end{figure*}

\begin{figure*}
\includegraphics[width=0.47\textwidth]{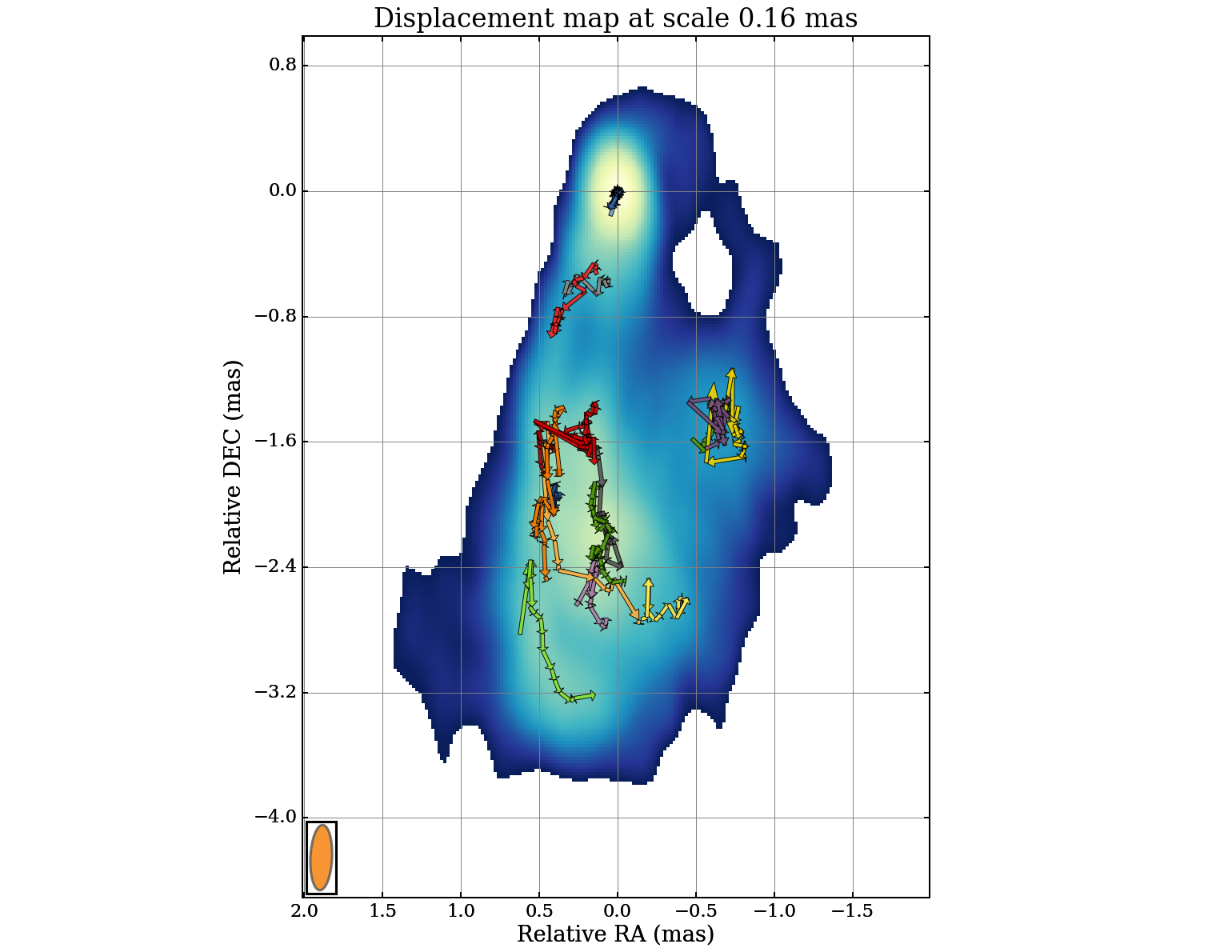}
\includegraphics[width=0.47\textwidth]{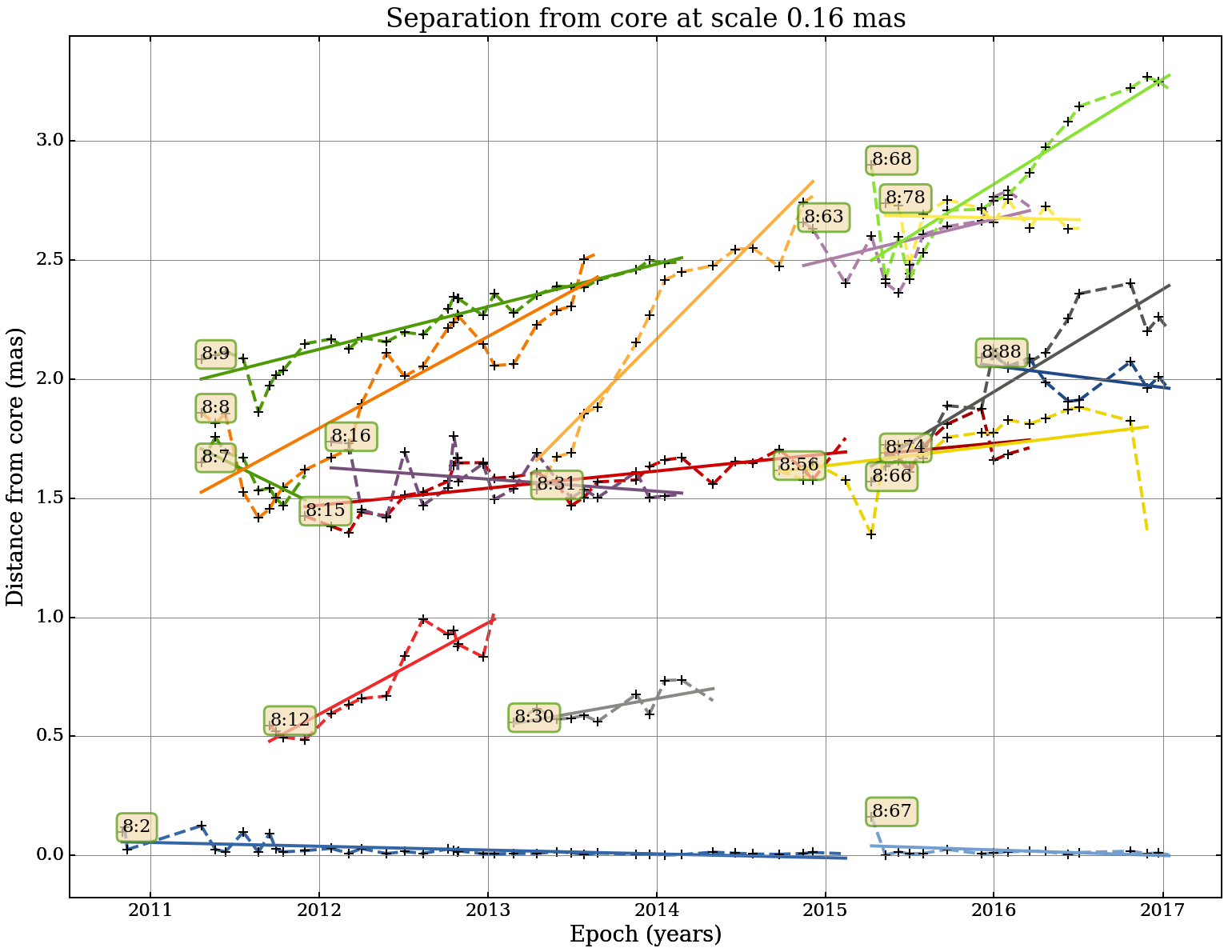}
\caption{\textbf{Projected kinematics (left) and fitted core-separation plots (right) with length scale 0.16\,mas, 8 component fit. } }\label{016mas8}
\end{figure*}

\begin{figure*}
\includegraphics[width=0.47\textwidth]{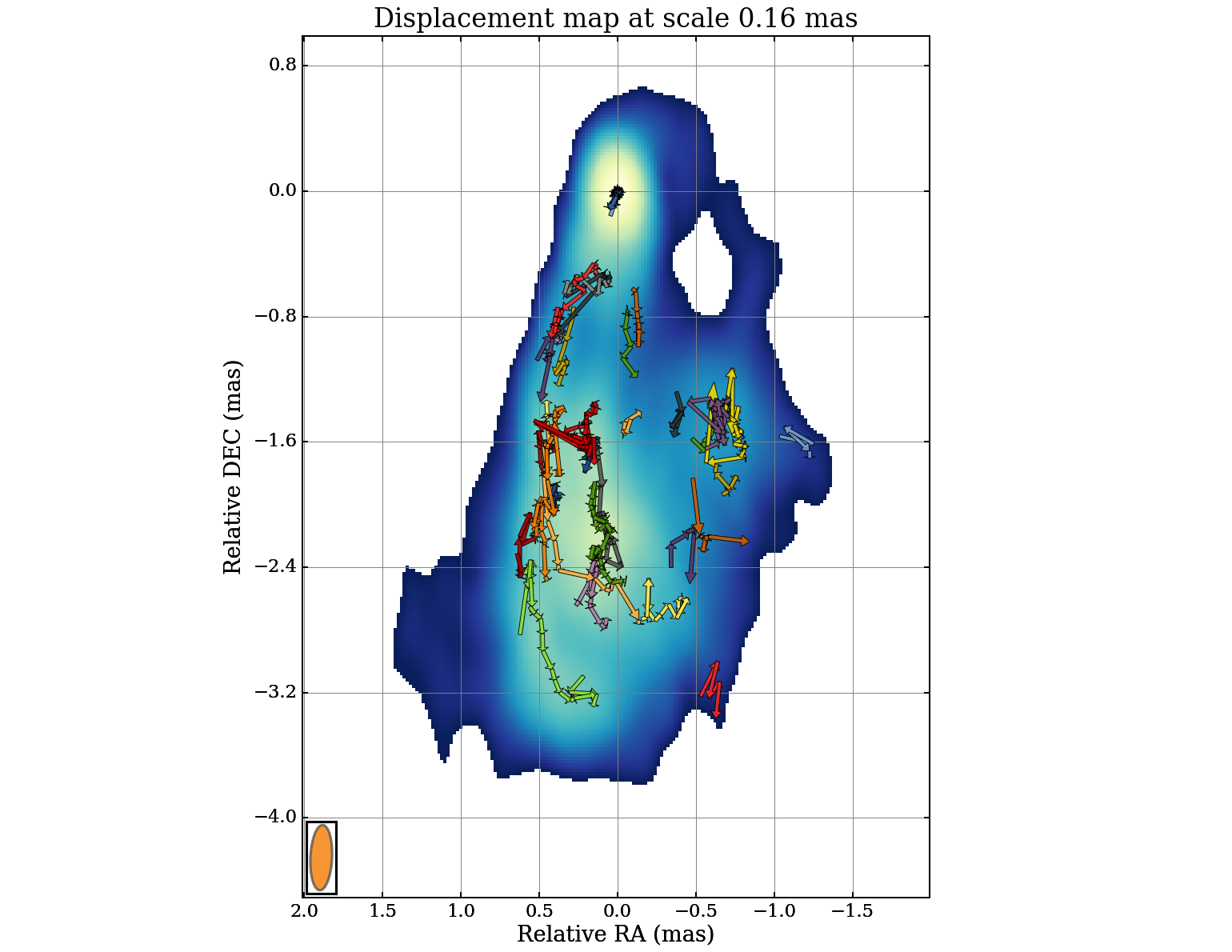}
\includegraphics[width=0.47\textwidth]{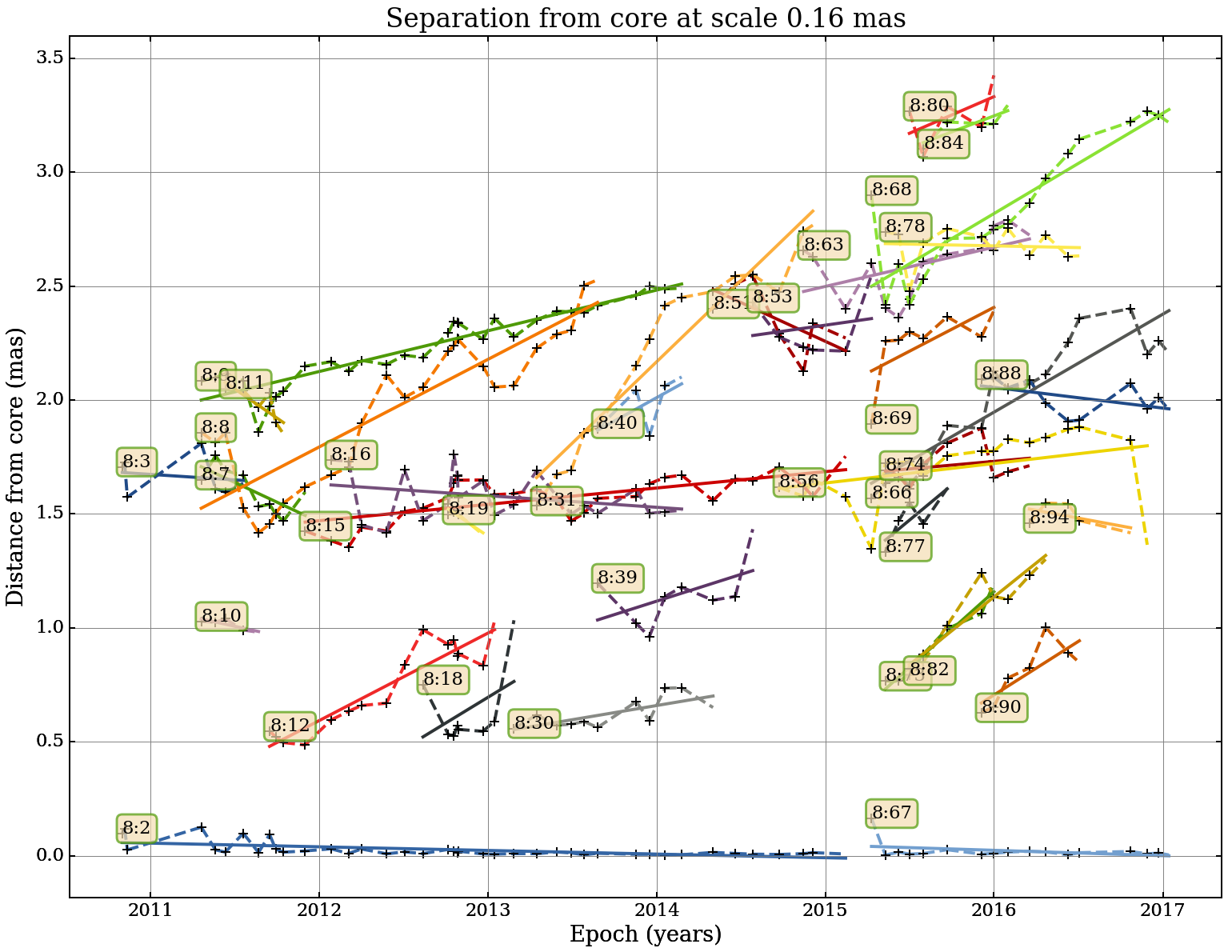}
\caption{\textbf{Projected kinematics (left) and fitted core-separation plots (right) with length scale 0.16\,mas, 4 component fit. } }\label{016mas4}
\end{figure*}

\begin{table}
	\centering
	\caption{WISE analysis results, 0.48\,mas, 8 components. The first column is the ID that the WISE algorithm automatically assigns to a component. The second column is the observed proper motion of the component. }
	\label{wise_048_8}
	\begin{tabular}{lc} 
		\hline
		ID & $\mu$  \\
		   &  [mas/yr]      \\
		\hline
         24:5 & 0.11 $\pm$ 0.04 \\
         24:3 & -0.09 $\pm$ 0.04 \\
         24:1 & 0.19 $\pm$ 0.01  \\
         24:9 & 0.52 $\pm$ 0.04  \\
         24:10 & 0.15 $\pm$ 0.02  \\
         24:7 & 0.04 $\pm$ 0.11  \\
         24:2 & -0.01 $\pm$ 0.00  \\

		\hline
	\end{tabular}
\end{table}

\begin{table}
	\centering
	\caption{WISE analysis results, 0.48\,mas, 4 components. Column descriptions the same as Table \ref{wise_048_8}.}
	\label{wise_048_4}
	\begin{tabular}{lc} 
		\hline
		ID & $\mu$  \\
		   &  [mas/yr]      \\
		\hline
         24:5 & 0.11 $\pm$ 0.04  \\
         24:3 & -0.09 $\pm$ 0.04  \\
         24:1 & 0.19 $\pm$ 0.01  \\
         24:9 & 0.52 $\pm$ 0.04  \\
         24:10 & 0.15 $\pm$ 0.02  \\
         24:7 & 0.04 $\pm$ 0.11  \\
         24:2 & -0.01 $\pm$ 0.00  \\

		\hline
	\end{tabular}
\end{table}



\begin{table}
	\centering
	\caption{WISE analysis results, 0.32\,mas, 8 components. Column descriptions the same as Table \ref{wise_048_8}.}
	\label{wise_032_8}
	\begin{tabular}{lc} 
		\hline
		ID & $\mu$  \\
		   &  [mas/yr]      \\
		\hline
         16:8 & 0.13 $\pm$ 0.04  \\
         16:17 & 0.38 $\pm$ 0.07  \\
         16:23 & -0.62 $\pm$ 0.21  \\
         16:3 & -0.01 $\pm$ 0.00  \\
         16:1 & 0.17 $\pm$ 0.01  \\
         16:15 & 0.15 $\pm$ 0.05  \\
         16:13 & 0.12 $\pm$ 0.05  \\
         16:4 & -0.09 $\pm$ 0.03  \\
         16:18 & 0.22 $\pm$ 0.02  \\

		\hline
	\end{tabular}
\end{table}





\begin{table}
	\centering
	\caption{WISE analysis results, 0.32\,mas, 4 components. Column descriptions the same as Table \ref{wise_048_8}. }
	\label{wise_032_4}
	\begin{tabular}{lc} 
		\hline
		ID & $\mu$  \\
		   &  [mas/yr]      \\
		\hline
         16:8 & 0.13 $\pm$ 0.04  \\
         16:17 & 0.38 $\pm$ 0.07  \\
         16:23 & -0.62 $\pm$ 0.21  \\
         16:3 & -0.01 $\pm$ 0.00  \\
         16:1 & 0.17 $\pm$ 0.01  \\
         16:15 & 0.15 $\pm$ 0.05  \\
         16:13 & 0.12 $\pm$ 0.05  \\
         16:4 & -0.09 $\pm$ 0.03  \\
         16:18 & 0.22 $\pm$ 0.02  \\

		\hline
	\end{tabular}
\end{table}



\begin{table}
	\centering
	\caption{WISE analysis results, 0.24\,mas, 8 components. Column descriptions the same as Table \ref{wise_048_8}.}
	\label{wise_024_8}
	\begin{tabular}{lc} 
		\hline
		ID & $\mu$  \\
		   &  [mas/yr]      \\
		\hline
         12:4 & 0.11 $\pm$ 0.06  \\
         12:61 & -0.03 $\pm$ 0.04  \\
         12:2 & -0.01 $\pm$ 0.00  \\
         12:47 & 0.44 $\pm$ 0.06  \\
         12:11 & 0.53 $\pm$ 0.03  \\
         12:36 & 0.13 $\pm$ 0.03  \\
         12:65 & -0.66 $\pm$ 0.20  \\
         12:15 & 0.17 $\pm$ 0.06  \\
         12:46 & 0.53 $\pm$ 0.08  \\
         12:1 & 0.27 $\pm$ 0.01  \\

		\hline
	\end{tabular}
\end{table}



\begin{table}
	\centering
	\caption{WISE analysis results, 0.24\,mas, 4 components. Column descriptions the same as Table \ref{wise_048_8}.}
	\label{wise_024_4}
	\begin{tabular}{lc} 
		\hline
		ID & $\mu$  \\
		   &  [mas/yr]      \\
		\hline
         12:15 & 0.17 $\pm$ 0.06  \\
         12:63 & 0.45 $\pm$ 0.08  \\
         12:11 & 0.53 $\pm$ 0.03  \\
         12:61 & -0.03 $\pm$ 0.04  \\
         12:2 & -0.01 $\pm$ 0.00  \\
         12:69 & -0.13 $\pm$ 0.17  \\
         12:12 & 0.11 $\pm$ 0.23  \\
         12:38 & 0.02 $\pm$ 0.22  \\
         12:24 & 0.09 $\pm$ 0.06  \\
         12:36 & 0.13 $\pm$ 0.03  \\
         12:56 & 0.05 $\pm$ 0.16  \\
         12:65 & -0.66 $\pm$ 0.20  \\
         12:55 & 0.35 $\pm$ 0.26  \\
         12:13 & 0.41 $\pm$ 0.13  \\
         12:53 & 0.06 $\pm$ 0.11  \\
         12:54 & 0.52 $\pm$ 0.08  \\
         12:25 & 0.28 $\pm$ 0.13  \\
         12:20 & -0.01 $\pm$ 0.08  \\
         12:4 & 0.11 $\pm$ 0.06  \\
         12:58 & 0.51 $\pm$ 0.10  \\
         12:47 & 0.44 $\pm$ 0.06  \\
         12:10 & 0.53 $\pm$ 0.10  \\
         12:46 & 0.53 $\pm$ 0.08  \\
         12:1 & 0.27 $\pm$ 0.01  \\
		\hline
	\end{tabular}
\end{table}



\begin{table}
	\centering
	\caption{WISE analysis results, 0.16\,mas, 12 components. Column descriptions the same as Table \ref{wise_048_8}.}
	\label{wise_016_12}
	\begin{tabular}{lc} 
		\hline
		ID & $\mu$  \\
		   &  [mas/yr]      \\
		\hline
         8:2 & -0.02 $\pm$ 0.00  \\
         8:15 & 0.07 $\pm$ 0.01  \\
         8:8 & 0.38 $\pm$ 0.04  \\
         8:66 & 0.43 $\pm$ 0.05  \\
         8:31 & 0.71 $\pm$ 0.06  \\
         8:67 & -0.02 $\pm$ 0.02  \\
         8:56 & 0.09 $\pm$ 0.05  \\
         8:16 & -0.05 $\pm$ 0.03  \\
         8:12 & 0.38 $\pm$ 0.04  \\
         8:9 & 0.18 $\pm$ 0.01  \\
         8:63 & 0.17 $\pm$ 0.09  \\
         8:68 & 0.44 $\pm$ 0.06  \\
		\hline
	\end{tabular}
\end{table}



\begin{table}
	\centering
	\caption{WISE analysis results, 0.16\,mas, 8 components. Column descriptions the same as Table \ref{wise_048_8}.}
	\label{wise_016_8}
	\begin{tabular}{lc} 
		\hline
		ID & $\mu$  \\
		   &  [mas/yr]      \\
		\hline
         8:78 & -0.02 $\pm$ 0.06  \\
         8:31 & 0.71 $\pm$ 0.06  \\
         8:15 & 0.07 $\pm$ 0.01  \\
         8:8 & 0.38 $\pm$ 0.04  \\
         8:30 & 0.12 $\pm$ 0.04  \\
         8:74 & 0.07 $\pm$ 0.10  \\
         8:7 & -0.35 $\pm$ 0.12  \\
         8:2 & -0.02 $\pm$ 0.00  \\
         8:67 & -0.02 $\pm$ 0.02  \\
         8:66 & 0.43 $\pm$ 0.05  \\
         8:63 & 0.17 $\pm$ 0.09  \\
         8:88 & -0.09 $\pm$ 0.05  \\
         8:68 & 0.44 $\pm$ 0.06  \\
         8:9 & 0.18 $\pm$ 0.01  \\
         8:16 & -0.05 $\pm$ 0.03  \\
         8:56 & 0.09 $\pm$ 0.05  \\
         8:12 & 0.38 $\pm$ 0.04  \\
		\hline
	\end{tabular}
\end{table}



\begin{table}
	\centering
	\caption{WISE analysis results, 0.16\,mas, 4 components. Column descriptions the same as Table \ref{wise_048_8}.}
	\label{wise_016_4}
	\begin{tabular}{lc} 
		\hline
		ID & $\mu$  \\
		   &  [mas/yr]      \\
		\hline
         8:51 & -0.34 $\pm$ 0.18  \\
         8:94 & -0.14 $\pm$ 0.12  \\
         8:40 & 0.41 $\pm$ 0.27  \\
         8:31 & 0.71 $\pm$ 0.06  \\
         8:8 & 0.38 $\pm$ 0.04  \\
         8:16 & -0.05 $\pm$ 0.03  \\
         8:10 & -0.16 $\pm$ 0.06  \\
         8:63 & 0.17 $\pm$ 0.09  \\
         8:67 & -0.02 $\pm$ 0.02  \\
         8:56 & 0.09 $\pm$ 0.05  \\
         8:90 & 0.47 $\pm$ 0.16  \\
         8:30 & 0.12 $\pm$ 0.04  \\
         8:77 & 0.61 $\pm$ 0.24  \\
         8:88 & -0.09 $\pm$ 0.05  \\
         8:53 & 0.11 $\pm$ 0.26  \\
         8:11 &-0.51 $\pm$ 0.19  \\
         8:3 & -0.05 $\pm$ 0.12  \\
         8:69 & 0.38 $\pm$ 0.18  \\
         8:12 & 0.38 $\pm$ 0.04  \\
         8:82 & 0.59 $\pm$ 0.09  \\
         8:78 & -0.02 $\pm$ 0.06  \\
         8:80 & 0.32 $\pm$ 0.30  \\
         8:2 & -0.02 $\pm$ 0.00  \\
         8:39 & 0.24 $\pm$ 0.16  \\
         8:66 & 0.43 $\pm$ 0.05  \\
         8:18 & 0.45 $\pm$ 0.37  \\
         8:74 & 0.07 $\pm$ 0.10  \\
         8:7 & -0.35 $\pm$ 0.12  \\
         8:73 & 0.66 $\pm$ 0.06  \\
         8:19 & -0.51 $\pm$ 0.13  \\
         8:68 & 0.44 $\pm$ 0.06  \\
         8:9 & 0.18 $\pm$ 0.01  \\
         8:84 & 0.28 $\pm$ 0.10  \\
         8:15 & 0.07 $\pm$ 0.01  \\
	\end{tabular}
\end{table}



\label{lastpage}
\end{document}